\documentclass[12pt,preprint]{emulateapj}

\usepackage{natbib}
\newcommand{\ngal}{\bar{n}_{\mathrm{g}}}

\newcommand{\Nsat}{N_\mathrm{sat}}
\newcommand{\Ncen}{N_\mathrm{cen}}

\newcommand{\hmpc}{h^{-1}\mathrm{Mpc}}
\newcommand{\hmpcVol}{h^{3}\mathrm{Mpc}^{-3}}
\newcommand{\hkpc}{h^{-1}\mathrm{kpc}}
\newcommand{\hMsun}{h^{-1}M_{\odot}}

\newcommand{\Mmin}{M_\mathrm{min}}
\newcommand{\Mstar}{M_{\ast}}
\newcommand{\Lstar}{L_{\ast}}

\newcommand{\Msub}{M_{\mathrm{sub}}}
\newcommand{\Mzero}{M_\mathrm{0}}
\newcommand{\Mone}{M_\mathrm{1}}

\newcommand{\Mhost}{M_\mathrm{host}}

\newcommand{\onehalo}{\mathrm{1h}}
\newcommand{\twohalo}{\mathrm{2h}}

\newcommand{\fsat}{f_\mathrm{sat}}

\newcommand{\xir}{\xi(r)}

\usepackage{epsfig}
\usepackage{graphicx} 
\usepackage{rotating}

\begin{document}

\title{A COSMIC COINCIDENCE: THE POWER-LAW GALAXY CORRELATION FUNCTION}

\author{Douglas~F.~Watson and Andreas~A.~Berlind\altaffilmark{1},}
\affiliation{Department of Physics and Astronomy, Vanderbilt University, Nashville, TN 37235}

\author{Andrew~R.~Zentner} \affiliation{Department of Physics and
Astronomy, The University of Pittsburgh, Pittsburgh, PA 15260}

\begin{abstract}
We model the evolution of galaxy clustering through cosmic time to
investigate the nature of the power-law shape of  $\xir$, the galaxy
two-point correlation function.  While $\xir$ on large scales is set
by primordial fluctuations,  departures from a power law are governed
by galaxy pair counts on small scales, subject to non-linear dynamics.
We assume that galaxies reside within dark matter halos and subhalos.
Therefore, the shape of  the correlation function on small scales
depends on the amount of halo substructure.   We use a semi-analytic
substructure evolution model to study subhalo populations within host
halos.   We find that tidal mass loss and, to a lesser extent,
dynamical friction dramatically deplete the  number of subhalos within
larger host halos over time, resulting in a $\sim 90\%$ reduction by
$z=0$ compared to  the number of distinct mergers that occur during
the assembly of a host halo.   We show that these non-linear processes
resulting in this depletion are essential for achieving a power-law
$\xir$.  We investigate how the shape of $\xir$ depends on subhalo
mass (or luminosity) and redshift.   We find that $\xir$ breaks from a
power law at high masses,  implying that only galaxies of luminosities
$ \lesssim \Lstar$ should exhibit power-law clustering.   Moreover, we
demonstrate that $\xir$ evolves from being far from a power law at
high redshift, toward a near  power-law shape at $z = 0$.  We argue
that $\xir$ will once again evolve away from a power law in the
future.   This is in large part caused by the evolving competition
between the accretion and destruction rates  of subhalos over time,
which happen to strike just the right balance at $z \approx 0$.   We
then investigate the conditions required for $\xir$ to be a power law
in a general context.   We use the halo model along with simple
parametrizations of the halo occupation distribution (HOD) to probe
galaxy occupation at various masses and redshifts.  We show that  key
ingredients determining the shape of $\xir$ are the fraction of
galaxies that are satellites,  the relative difference in mass between
the halos of isolated galaxies and halos that contain a single
satellite on average, and the rareness of halos that host galaxies.
These pieces are intertwined and we  find no simple, universal rule
for which a power-law $\xir$ will occur.  However, we do show that the
physics responsible for setting the galaxy content of halos do not
care about the conditions needed  to achieve a power law $\xir$ and
these conditions are met only in a narrow mass and redshift range.  We
conclude that the power-law nature of $\xir$ for $\Lstar$ and fainter
galaxy samples at low redshift  is a cosmic coincidence.
\end{abstract}

\altaffiltext{1}{Alfred P. Sloan Fellow}

\section{INTRODUCTION}
\label{intro}

The two-point correlation function of galaxies was measured four
decades ago and found to be consistent with a $\xir \propto r^{-2}$
power law
\citep{totsuji69,peebles73,hauserpeebles73,peebleshauser74,peebles74}.
Since that time, successively larger galaxy redshift surveys
\citep[e.g.,][]{huchra83,daCosta88,santiago95,
shectman96,saunders00,colless01,york00a} have mapped the distribution
of galaxies with ever increasing  precision and confirmed correlation
functions consistent with power laws over a large range of scales
\citep[e.g.,][]{delapparent88,marzke95,hermit96,tucker97,jing98,jing02,norberg02,zehavi02}.
The scales on which  a single power-law description is valid span a
range from large regions exhibiting mild density fluctuations  ($r
\gtrsim 10$~Mpc), to smaller regions with large density fluctuations
experiencing rapid non-linear evolution  ($r \sim 1-10$~Mpc), to
collapsed and virialized galaxy groups and clusters ($r \lesssim
1$~Mpc).  It has long  been noted that the lack of any feature
delineating the transitions among these scales is surprising
\citep[e.g.,][]{peebles74,gott_turner79,hamilton_tegmark02,masjedi06a,liwhite09}.
This is especially true given  that the matter correlation function in
the now well-established concordance cosmological model differs
significantly from a power law.  In this paper, we return to this
long-standing problem and address the origin  of a power-law galaxy
correlation function in the context of our modern paradigm for the
growth of cosmic structure.

This conundrum can be refined within the contemporary framework in
which galaxies live within virialized halos of dark matter
\citep{whiterees78,blumenthal_etal84}.  In such a model, galaxy
clustering statistics can be modeled  as a combination of dark matter
halo properties and a halo occupation distribution (HOD) that
specifies how galaxies  occupy their host halos
\citep[e.g.,][]{peacock00a,scoccimarro01a,berlind02,cooray02}.  In
this {\em halo model}  approach, the galaxy correlation function is a
sum of two terms: On small scales, pairs of galaxies reside in the
same host dark matter halo (the ``one-halo'' term), whereas on large
scales, the individual galaxies of a pair  reside in distinct halos
(the ``two-halo'' term).  These two terms depend on the HOD in
different ways, requiring  delicate tuning in order to spawn an
unbroken power law \citep[e.g.,][]{berlind02}.  Consequently, a
feature in  $\xir$ at scales corresponding to the radii of the
typical, viralized halos that host luminous galaxies is expected.

In a dramatic success for the halo model, \citet{zehavi04a} first
detected a statistically-significant departure from a power law due to
the high precision measurements of the Sloan Digital Sky Survey, and
demonstrated that the halo model provides an acceptable fit to the
data.  \citet{zehavi05a}  confirmed this result, adding that power-law
departures grow stronger with galaxy luminosity \citep[see
also][]{blake08,ross10}. $\xir$ has since been shown to deviate from a
power law at high redshifts \citep{ouchi05,lee06,coil06,wake_etal11}.
Nevertheless, it remains a  fact that deviations from a power law at
low redshifts are small and the galaxy correlation function is roughly
a power law over an enormous range of galaxy-galaxy separations.
Deviations have been revealed only through ambitious observational
efforts.

Halos are known to be replete with self-bound structures, dubbed
``subhalos'' \citep{Ghigna98,Klypin99a,moore_etal99},  and both halos
and subhalos are thought to be the natural sites of galaxy formation.
Subhalos were isolated halos in  their own right, hosting distinct
galaxies before merging into a larger group or cluster
halo\footnote[1]{\emph{Satellites} or \emph{subhalos} are used
throughout the paper to refer to self-bound entities lying  within the
virial radius of a larger halo.  Those that do not lie within a larger
system are designated as  \emph{centrals}, \emph{host halos} or simply
\emph{hosts}.}.  Remarkably, the clustering of host halos along with
their  associated subhalos is very similar to that of observed
galaxies \citep{kravtsovklypin99,colin99,kravtsov04a},  suggesting a
simple correspondence of galaxies with host halos and subhalos.  This
was clearly demonstrated by  \citet{conroy06} who compared the
correlation functions of hosts and subhalos to that of galaxies over a
broad range of  luminosities and redshifts ($z \sim 0-4$), finding
excellent agreement.  These results indicate that an understanding of
the physics governing the subhalo populations within host halos may
provide insight into the physics of galaxy clustering  and the near
power-law form of the galaxy two-point correlation function.

In this paper, we examine the causes of the observed power-law
correlation function by studying the mergers,  survival, and/or
destruction of dark matter subhalos.  Our focus in this paper is on
the gross features of the galaxy  two-point function and {\em not} on
detailed comparisons to specific data sets.  We explore more
sophisticated galaxy-halo assignments and statistical comparisons with
data in a  forthcoming follow-up study (Watson et al. in prep.).

We argue that the nearly power-law, low-redshift galaxy correlation
function is a coincidence.   The correlation function of common $L
\lesssim \Lstar$ galaxies evolves from relatively  strong small-scale
clustering at early times, through a power-law at the present epoch,
and most likely  toward relatively weak small-scale clustering in the
future.  The origin of the present-day power law, in turn, relies  on
the tuning of several disconnected ingredients, at least three of
which are: the normalization of primordial density  fluctuations
determined by early Universe physics; a halo mass scale for efficient
galaxy formation determined largely  by atomic physics, stellar
physics, and the physics of compact objects; and relative abundances
of baryonic matter,  dark matter, and dark energy in the Universe.

Our paper is organized as follows.   In \S~\ref{halomodel} we review
the halo model and restate the problem in terms of this framework.  In
\S~\ref{model} we give an overview of our primary modeling technique.
In \S~\ref{dynamics} we investigate the individual roles of merging,
dynamical friction, and mass loss in shaping the  halo occupation
statistics of subhalos, as well as the resulting halo correlation
function.   In \S~\ref{dependence} we show how $\xir$ depends on host
halo mass and redshift.   In \S~\ref{make_powerlaw} we explore a
standard parametrization of the HOD to see what is required to get a
power-law  $\xir$, and we predict the masses and redshifts at which a
power-law $\xir$ can be constructed.   In \S~\ref{summary} we give a
summary of our results and our primary conclusions.   Throughout this
paper, we work within the standard, vacuum-dominated, cold dark matter
($\Lambda$CDM) cosmological  model with $\Omega_{\mathrm{m}}=0.3$,
$\Omega_{\Lambda}=0.7$, $\Omega_{\mathrm{b}}=0.04$, $h_{0}=0.7$,
$\sigma_{8}=0.9$,  and $n_{\mathrm{s}}=1.0$.  These values differ
slightly form the WMAP best-fit values, however this has little effect
on  our general results and was chosen in order to compare to previous
work that used similar cosmological models.

\section{A Modern Restatement of the Problem in Halo Model Language}
\label{halomodel}

\begin{figure}
\begin{center}
\includegraphics[width=.5\textwidth]{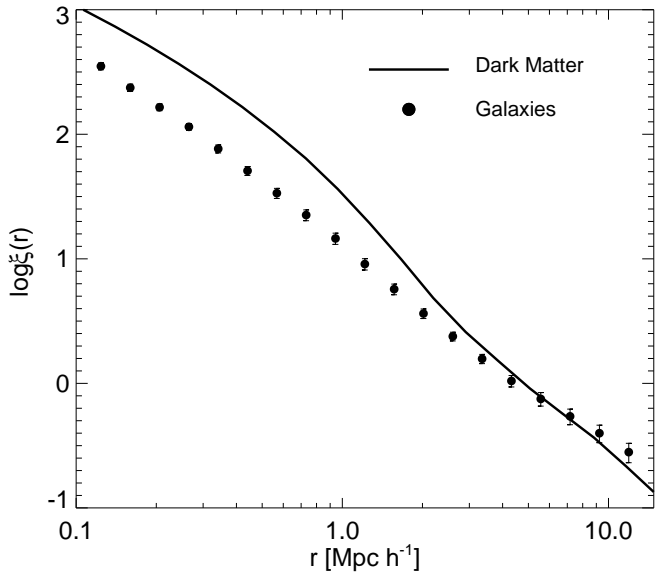}
\caption{Correlation function of galaxies compared to dark matter.
Points show the correlation function of galaxies from the APM survey,
estimated from deprojecting the angular correlation function
\citep{maddox90,baugh96}.  The curve shows the correlation function of
dark matter measured from the LCDM GIF simulation run by the Virgo
collaboration \citep{jenkins98}.}
\label{fig:ximass}
\end{center}
\end{figure}

Though the observed galaxy correlation function is nearly a power law,
the matter correlation function predicted by  the concordance
cosmological model is not.  This is evident in
Figure~\ref{fig:ximass}.  On scales corresponding to  collapsed
objects, the dark matter correlation function exceeds the values that
would be obtained by extrapolating the  larger-scale power law to
small scales.  However, galaxies are biased with respect to dark
matter in such a way as to  counteract this excess.  We can examine
this discrepancy in terms of the halo model.  If the reader is familiar
with the halo model formalism, he or she may wish to skip to
\S~\ref{thebattle}

\subsection{Halo Model Basics}
\label{halomodel}

Assuming that all galaxies live within virialized dark matter halos,
the galaxies comprising any  pair can come either from within the same
halo (the {\em one-halo term}) or from two separate  halos (the {\em
two-halo term}).  The correlation function is then given as the sum of
these two terms
\begin{equation}
\xi(r)=\xi(r)^{\onehalo}+\xi(r)^{\twohalo} + 1,
\end{equation}
(e.g., \citealt{cooray02}; for this particular form of the equation
see \citealt{zheng04a}).  The probability  distribution $P(N|M)$ that
a halo of mass $M$ contains $N$ galaxies together with the spatial
distribution of  galaxies within their host halos constitute the {\em
halo occupation distribution} (HOD).  We denote the first and  second
moments of $P(N|M)$ at a specific mass $M$ as $\langle N\rangle_{M}$
and $\langle N(N-1)\rangle_{M}$,  respectively.
The one-halo term can be computed by counting the average number of
galaxy pairs of a given separation in a common  halo and averaging
over all halos.  We write the one-halo term as \citep{berlind02}
\begin{eqnarray}
\label{eqn:xi_1h}
1 + \xi(r)^{\onehalo} = \frac{1}{2\pi r^{2}\ngal^2}\int
dM\frac{dn}{dM} \\ \times \frac{\langle N(N-1)\rangle_{M}}{2} F(r|M)
\nonumber
\end{eqnarray}
where $dn/dM$ is the halo mass function, $\langle N(N-1)\rangle_{M}/2$
is the mean number of galaxy pairs within a  halo of mass $M$, and
$F(r|M)$ is the distribution of separations between these
pairs\footnote{This notation is  slightly different from that used in
\citet{berlind02}, in which $F(r)$ denoted the {\em cumulative}  pair
distribution.}.  If the average spatial distribution of galaxies
within their host halos is  $\lambda(r|M)$, then the pair separation
distribution $F(r|M)$ is the convolution of  $\lambda(r|M)$ with
itself.  The quantity $\ngal$ is the mean density of galaxies in the
Universe,
\begin{equation}
\label{eqn:ngal}
\ngal = \int dM\frac{dn}{dM}\langle N \rangle _{M}.
\end{equation}

Motivated by theoretical considerations
\citep[e.g.,][]{berlind03,kravtsov04a,zheng05}, the HOD of galaxies is
usually  considered separately for central galaxies that live near the
centers of their host halos and satellite galaxies that  orbit within
the host halo potential.  Each halo above some mass threshold should
contain one central galaxy and  possibly one or more satellites,
depending on the host mass and the HOD.  In this framework it is
useful to consider  contributions to the one-halo term separately for
central-satellite and satellite-satellite pairs.  Therefore, we
rewrite the one-halo term as \citep{berlind02}
\begin{eqnarray}
\label{eqn:xi_1h_cs}
1 + \xi(r)^{\onehalo} & = & \frac{1}{2\pi r^{2}\ngal^2}\int
dM\frac{dn}{dM} \\ \times \ \Big[\langle \Ncen \Nsat \rangle _{M}
F_{\mathrm{cs}}(r|M) & + & \frac{\langle \Nsat(\Nsat-1) \rangle
_{M}}{2} F_{\mathrm{ss}}(r|M) \Big],\nonumber
\end{eqnarray}
where $\langle \Ncen \Nsat \rangle _{M}$ and $\langle \Nsat(\Nsat-1)
\rangle _{M}/2$ are the mean number of  central-satellite and
satellite-satellite pairs in hosts of mass $M$, and
$F_{\mathrm{cs}}(r|M)$ and  $F_{\mathrm{ss}}(r|M)$ are the pair
separation distributions of central-satellite and satellite-satellite
pairs,  respectively.  If the central galaxies always reside very
close to the center of the host halo and the average  distribution of
satellite positions within the host halo is
$\lambda_{\mathrm{s}}(r|M)$, then
$F_{\mathrm{cs}}=\lambda_{\mathrm{s}}(r|M)$ and $F_{\mathrm{ss}}(r|M)$
is the convolution of $\lambda_{\mathrm{s}}(r|M)$  with itself.  In
practical cases there is at most one central galaxy and satellites are
only present in halos with a  central, so that $\langle \Ncen \Nsat
\rangle_{M} = \langle \Nsat \rangle_{M}$.  The total fraction of
galaxies that  are satellites in a sample is then
\begin{eqnarray}
\label{eqn:fsat}
\fsat & = & \ngal^{-1}\, \int\, dM\, \frac{dn}{dM} \langle \Nsat
      \rangle _{M} \nonumber \\ & = & \frac{\int\, dM\, \frac{dn}{dM}
      \langle \Nsat \rangle _{M}}{\int\, dM\, \frac{dn}{dM}  (\langle
      \Ncen \rangle _{M} + \langle \Nsat \rangle _{M})}.
\end{eqnarray}
The satellite fraction, $\fsat$, will prove an important quantity in
determining the  shape of the galaxy correlation function.

On scales significantly larger than individual halos, the two-halo
term dominates the clustering strength.  It is  most simply written in
Fourier space as (\citealt{cooray02}; for this particular form of the
equation see  \citealt{tinker05})
\begin{eqnarray}
\label{eqn:xi_2h}
P^{\twohalo}(k) = P_\mathrm{m}(k)\Bigg[ \ngal^{-1} \int dM
\frac{dn}{dM} \langle N \rangle_{M} \\ \times \ b_{h}(M,r)
\tilde{\lambda}(k|M)\Bigg]^{2}, \nonumber
\end{eqnarray}
where $P_\mathrm{m}(k)$ is the matter power spectrum, $b_{h}(M,r)$ is
a (possibly scale-dependent)  halo bias function, and
$\tilde{\lambda}(k|M)$ is the Fourier transform of the spatial number
density of galaxies  within their host halos.  We can invert the
Fourier transform of the two-halo power spectrum to recover the
two-halo  term of the correlation function.  In the limit that the
galaxy pair separation is larger than any halo of interest,  the
two-halo term becomes
\begin{eqnarray}
\label{eqn:xi_2h_lim}
\xi^{\twohalo}(r) & \simeq & \left[ \ngal^{-1} \int \, b_{h}(M,r)\,
                 \langle N\rangle_{M} \frac{dn}{dM}\, \mathrm{d}M
                 \right]^2 \xi_{m}, \\ &   =    & b_{\mathrm{g}}^2\,\
                 \xi_{m}, \nonumber
\end{eqnarray}
where $\xi_{m}(r)$ is the matter correlation function.
Equation~(\ref{eqn:xi_2h_lim}) explicitly shows that the  large-scale
galaxy correlation function is essentially the halo correlation
function except halos of different masses  are weighted by $\langle N
\rangle_{M}$.    The galaxy bias describing the relative clustering of
galaxies to dark matter  $b_{\mathrm{g}} = \sqrt{\xi/\xi_{m}}$ is the
quantity in square brackets in Equation~(\ref{eqn:xi_2h_lim}).

\subsection{The Battle of the 1- Halo and 2- Halo Terms}
\label{thebattle}

\citet{berlind02} showed that maintaining a power-law correlation
function requires a careful balance between the  one-halo and two-halo
terms and is thus quite difficult to achieve.  This is because the
one-halo term generally  changes by a larger amount than the two-halo
term in response to changes to the HOD.  A close examination of
Equations~(\ref{eqn:ngal}),~(\ref{eqn:xi_1h_cs}),~(\ref{eqn:xi_2h})
and~(\ref{eqn:xi_2h_lim}) reveals why this is the  case.

Consider first the two-halo term as it is the simplest.  On large
scales, the two-halo term is just a weighted average  of the
clustering of host halos.  For simplicity, assume (albeit incorrectly)
the halo bias to be a constant function  of halo mass.  Increasing
$\langle N \rangle_M$ increases both the number of two-halo pairs at a
given separation  (the square of the integral in
Eqs.~[\ref{eqn:xi_2h}] and~[\ref{eqn:xi_2h_lim}]) and the number of
random pairs  $\ngal^2/2$, by the same amount.  The reason the
two-halo term is at all sensitive to the HOD is that the bias of
halos does depend on mass and so changing the relative number of
galaxies in high-mass vs. low-mass halos changes the  weight in the
average of the halo bias in Equation~(\ref{eqn:xi_2h_lim}).  For
example, assigning a large number of  satellite galaxies to high-mass
halos increases $\xi^{\twohalo}(r)$  by weighting highly-biased,
high-mass halos more heavily.   The possible range in the amplitude of
the two-halo term is limited by the variation of the halo bias
function $b_h(M)$,  within the mass range relevant to galaxies,
$10^{11} \lesssim M/M_{\odot} \lesssim 10^{15}$.  At low masses, the
halo  bias is $b_{h} \sim 0.65$ while, in the cluster regime ($M \sim
10^{14}\hMsun$), it grows to values of $b_{h} \sim 2$
\citep{tinker05}.  Bias continues to grow with mass, but more massive
halos are rare and do not contribute much to the  weighted average
because $dn/dM$ is miniscule.  The two-halo term scales like the
square of  the average bias $b_{\mathrm{g}}$ in
Equation~(\ref{eqn:xi_2h_lim}), so  the possible dynamic range
$\xi^{\twohalo}$ can display is, at most, a factor of $\sim 9$  and is
usually significantly smaller.  Simply put, the two-halo term depends
weakly on the HOD  because on large scales it is not possible to make
galaxies significantly more or less clustered  than the host halos
they occupy.

On small scales, the one-halo term dominates and the situation is
different.  The number of galaxy pairs within  an individual halo
scales with $\langle N(N-1)\rangle _{M}$ while the number of random
pairs scales with $\ngal^2$,  or $\langle N\rangle_{M}^2$.  It is
instructive to break the HOD into central and satellite galaxies.   In
the regime where there is one central galaxy per halo, the mean number
of central-satellite pairs is  $\langle \Ncen \Nsat \rangle_{M} =
\langle\Nsat\rangle_{M}$, whereas the mean number of
satellite-satellite pairs is  $\langle \Nsat(\Nsat-1) \rangle_{M}/2$.
Assuming a Poisson distribution for the number of satellite galaxies
\citep{kravtsov04a}, $\langle \Nsat (\Nsat - 1) \rangle_{M} = \langle
\Nsat \rangle_{M}^2$.  The mean number of random  pairs scales like
$(1+\langle\Nsat\rangle)_{M}^2$.  In the limit $\langle \Nsat
\rangle_{M} \gg 1$, the number of  satellite-satellite pairs dominates
the number of central-satellite pairs, but in this limit both the
number of  one-halo pairs and the square of the mean galaxy number
density scale as $\langle \Nsat \rangle_{M}^2$ so the one-halo  term
saturates to a maximum value and is insensitive to the number of
satellite galaxies per halo.

In most practical cases, the fraction of satellite galaxies in an
observational sample is $\fsat \lesssim 0.25$, so  samples tend to be
dominated by halos with satellite galaxy populations in the opposite
limit,  $\langle\Nsat\rangle_{M} \ll 1$.  This is due to the fact that
very massive host halos are rare, so halos with  $\langle \Nsat
\rangle_{M} > 1$ are rare.  With $\langle \Nsat \rangle_{M} \ll 1$,
the central-satellite term dominates and the number of such pairs
scales as $\langle \Nsat \rangle_{M}$  while the mean number density
$\ngal$ is approximately constant.  Examination of
Equations~(\ref{eqn:xi_1h_cs})  and~(\ref{eqn:fsat}) reveals that in
this regime $\xi^{\onehalo}$ scales {\em in proportion to the fraction
of satellite  galaxies and in inverse proportion to the number of host
halos}.  Host halo mass is largely fixed by requiring the  galaxies in
any sample to have an appropriate average number density (this is why
rare galaxies exhibit strong  small-scale clustering).  Therefore, the
one-halo term describing any given sample varies approximately
linearly with  $\langle \Nsat \rangle_{M}$ until $\langle \Nsat
\rangle > 1$, at which point it saturates.  It is interesting  that
nearly all the sensitivity of the correlation function to the HOD
comes from central-satellite galaxy  pairs in host halos where
satellite galaxies are uncommon!

In this work, we aim to understand the origin of the nearly power-law
galaxy correlation function.  The relevant  question is why is it that
the number of galaxies (or satellite galaxies to be more specific) per
halo is set just so  that the one-halo and two-halo terms in the
galaxy correlation function match smoothly, leaving only small
deviations  from a single power law over several orders of magnitude
in scale?  We confront this problem by studying the  properties and
evolution of subhalo populations.  We now turn to some of the details
of our modeling  methods.

\section{Overview of Halo Substructure Modeling}
\label{model}

Our approach is to study the evolution of subhalos within virialized
host halos as a method to understand satellite  galaxies and, in turn,
the evolution of galaxy clustering.  We focus our attention on the
relative strengths of  small-scale and large-scale clustering.  We
study subhalo populations using the approximate semi-analytic model of
\citet[][hereafter Z05]{zentner05}.  In this section, we briefly
review the fundamental aspects of the model that are  of immediate
relevance and we refer the reader to Z05 for details and validation.
The subhalo model is based on  \citet{zentner03} and is similar to the
independent models of \citet{TB04a,TB05b,TB05c} and
\citet{PenarrubiaBenson05},  while sharing many features with other
approximate treatments of halo substructure
\citep{OguriLee04,vdBosch05,faltenbacher_matthews05,purcell_etal07,giocoli_etal08,giocoli_etal09}.

Semi-analytic models are an approximation to the calculations of large
$N$-body simulations, yet such models offer many  advantages: (1)
semi-analytic calculations are computationally inexpensive; (2) they
have no inherent resolution  limits; (3) they enable the statistical
study of subhalos within very large numbers of host halos; (4) they
allow the  growth and mass-loss histories of particular subhalos to be
tracked without significant post-processing and analysis;  (5) they
make studies of model parameter space tractable; and (6) semi-analytic
models facilitate parsing complex  physical phenomena so that the
relative importance of different physical effects may be understood.
Our goal is to  quantify the relative importance of merging, which
increases subhalo abundances, and dynamical friction and mass loss,
which decrease subhalo abundances.  We also aim to explore predictions
for subhalo populations and galaxy correlation  functions from high
redshift to several Hubble times in the future.  Z05 extensively
tested the model we use in this  paper and showed that the model
produces subhalo mass functions, occupation statistics, and radial
distributions within  hosts that are in good agreement with a number
of high-resolution $N$-body simulations \citep[see the recent
comparison  in][as well]{koushiappas_etal10}.

The analytic model proceeds in several steps.  For a host halo of a
given mass $M$, observed at a given redshift $z$,  we generate a halo
merger tree using the mass-conserving implementation of the excursion
set formalism  \citep{bond91,LaceyCole93,LaceyCole94} developed by
\citet[][see \citealt{zentner07} for a review]{somerville99}.   This
yields a complete history of the masses and redshifts of all halos
that merged to form the final, target halo of  mass $M$ at redshift
$z$.  The host halo is the largest halo at each point in the merger
tree.  We model the density  distributions of all halos as
\citet[][hereafter NFW]{nfw97} profiles with concentrations determined
by their merger  histories according to \citet{wechsler02}.  At the
time of each merger, we assign the subhalo initial orbital parameters
drawn from distributions measured in N-body simulations (Z05, see
\citealt{benson05} for similar formalisms).  We then  integrate each
subhalo orbit within the host halo gravitational field, taking into
account dynamical friction and mass  loss.  We estimate dynamical
friction with an updated form of the \citet{chandrasekhar43}
approximation  \citep{hashimoto_etal03,zentner03}, account for
internal heating so that scaling relations describing the internal
structures of subhalos are obeyed
\citep{hayashi03,kazantzidis_etal04,kravtsov04b}, and allow for loss
of material  beyond the tidal radius on a timescale comparable to the
local dynamical time.  The details of each ingredient are  given in
Z05.

The correlation function of halos and subhalos and their associated
galaxies is sensitive to the abundance of subhalos  that survive both
possible mergers with the central, host galaxy due to dynamical
friction as well as mass loss and thus  remain as distinct objects in
orbit within their host halos with their galaxies intact.  Therefore,
it is necessary to  specify conditions under which the galaxy within a
subhalo may be ``destroyed'' and removed from our samples.  In this
work, we consider the clustering of mass-threshold samples of halos
and subhalos as a proxy for luminosity-threshold  samples of galaxies,
so significant mass loss will lead to a galaxy that is either
destroyed or dropped out of our  sample.  We assume such a scaling
between halo mass and galaxy luminosity {\em solely for the sake of
simplicity}.   Our primary points are qualitative in nature, but we
note that this is similar to other schemes that have described  data
successfully
\citep[e.g.,][]{kravtsovklypin99,colin99,kravtsov04a,tasitsiomi_etal04,conroy06}
and our calculations  with similar, but more sophisticated assignments
do not alter any of our basic results or conclusions.  In rare cases,
subhalos may survive close encounters with the center of their host
halo potentials.  We remove all subhalos that have  orbital apocenters
$r_{\rm apo} < 5$~kpc.  This choice is physically motivated because
the galaxies within such subhalos  would likely have merged with the
central galaxy, or at least be observationally indistinguishable from
the central  galaxy.  This choice is relatively conservative in that
galaxies on larger orbits would also likely be influenced and it  only
affects the results of calculations in which tidal mass loss is not
permitted (see below).  The net result of  evolving orbits for each
subhalo in the merger tree is a catalog of all surviving subhalos in
the final host halo at the  time of observation.  In some cases, a
halo that merges into a larger host contains subhalos of its own.
These  subs-of-subs are only abundant inside very large host masses
and are present in our model.

One of our aims is to study the individual roles of halo merging,
dynamical friction, and mass loss on the clustering  of halos.
Therefore, we compute subhalo populations in four different sets of
circumstances:
\begin{description}
\item[\emph{\textbf{No Effects}}] - a ``bare-bones'' model that does
not allow satellite galaxies to be modified by  dynamical friction or
mass loss.  In this case, any infalling subhalo remains intact, and we
assume that this subhalo  harbors a galaxy that will survive forever.
This is tantamount to assuming that galaxies form in all
sufficiently-large peaks in the primordial density field and survive
until today.
\item[\emph{\textbf{Fric. Only}}] - a model that only considers the
effects of halo merging and dynamical friction.   Subhalos never lose
mass and can only be destroyed by sinking to the very centers of their
hosts.
\item[\emph{\textbf{Strip. Only}}] - a model that only considers halo
merging and mass loss and assumes no dynamical  friction or central
merging.  Subhalos can lose mass and drop out of a mass threshold
sample, but they cannot lose  orbital energy and sink to the center of
the host potential.
\item[\emph{\textbf{Full}}] - our full model treating halo merging,
dynamical friction, and mass loss.  This is the  model that was
developed in Z05 and validated against $N$-body simulations.
\end{description}
We run our models for host masses\footnote{We note that we use the
``virial'' definition of a halo in which a halo is  defined as a
spherical region of mean density equal to $\Delta_\mathrm{vir}$ times
the mean background density.  For  our cosmological model,
$\Delta_\mathrm{vir}=337$ at $z=0$ and approaches 178 at high $z$.} in
the range from  $\mathrm{log}(\Mhost/\hMsun) = 11.0$ to $15.0$ in
steps of $\Delta(\mathrm{log}\Mhost) = 0.1$.  For each of these
masses, we run $1000$ statistical model realizations representing
different realizations of the local density field  and different halo
merger histories.  In this way, we sample the statistical properties
of subhalo populations over  the entire range of host halo masses
relevant to galaxy-galaxy correlations.  We repeat this process for
host masses  at $z=0$ as well as two past redshifts, $z=3$ and $z=1$,
and two future redshifts, $z=-0.6$, and $z=-0.9$.

\section{EFFECTS OF SUBHALO DYNAMICS ON THE GALAXY CORRELATION FUNCTION}
\label{dynamics}

\subsection{Halo Occupation Distribution Statistics}
\label{HOD}

The galaxy correlation function may be considered primarily a function
of the galaxy HOD \citep[e.g.,][]{berlind02}.   The prevailing
cosmological model is now stringently constrained and may be
considered fixed for our  purposes.  Moreover, theoretical predictions
of the abundances, clustering, and structures of host dark matter
halos in  the concordance cosmology are now well established.
Consequently, we focus on the properties of the HOD and the manner  in
which the HOD determines galaxy clustering.

We expect that each host halo of sufficient size contains one
dominant, central galaxy associated with the host itself,  as well as
additional satellite galaxies that are associated with relatively
large subhalos.  Thus, the HOD of galaxies  should resemble the HOD of
all halos (hosts plus their subhalos), and such a model is bolstered
by significant empirical  support
\citep{kravtsovklypin99,colin99,kravtsov04a,tasitsiomi_etal04,conroy06}.
As a result, we concentrate on the  insight that can be gleaned about
the development of the HOD of all halos, paying particular attention
to the separate  effects of halo mergers, dynamical friction, and mass
loss.

\begin{figure*}
\begin{center}
\includegraphics[width=\textwidth]{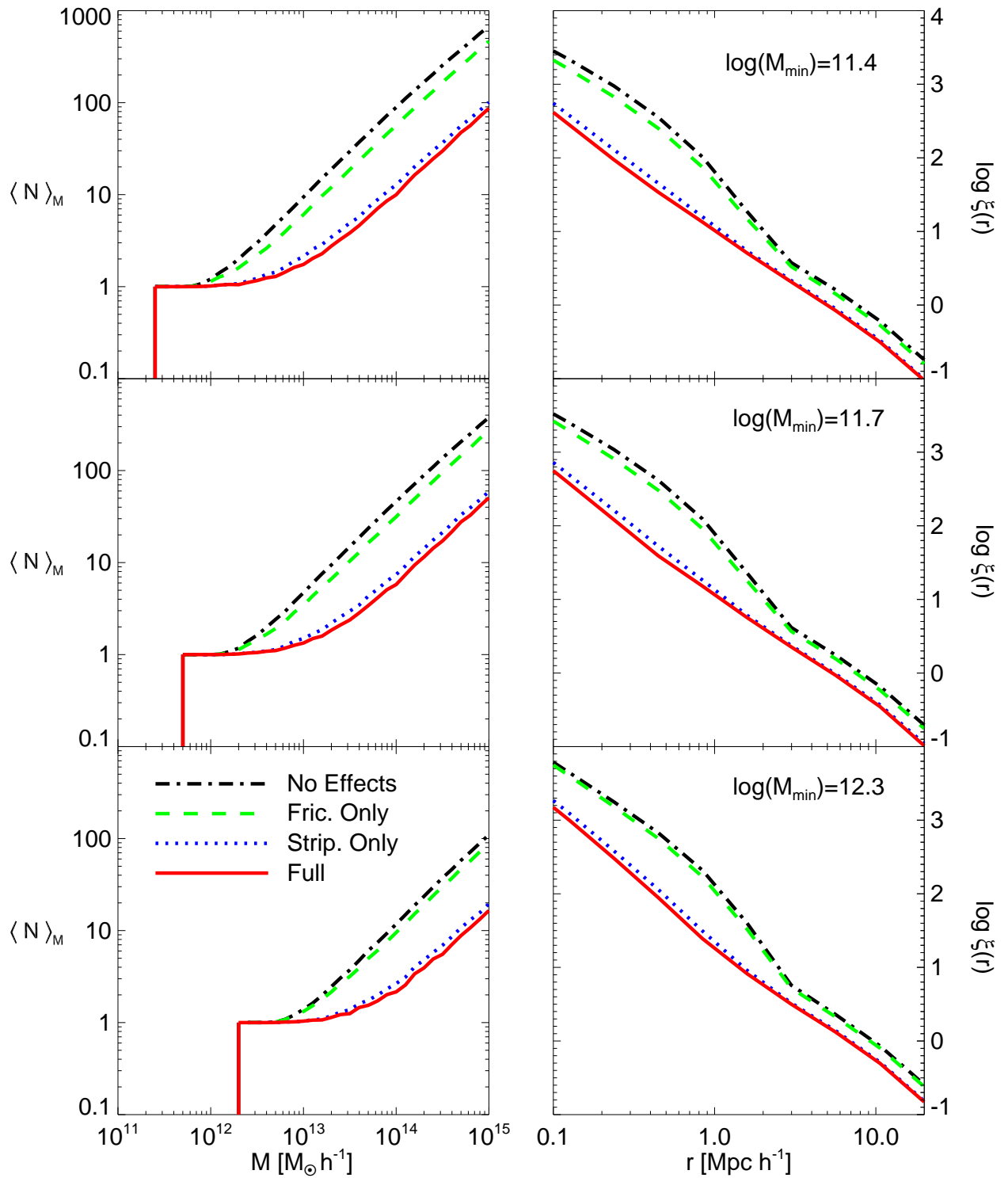}
\caption{  \emph{Left Panels}: Mean number of all halos (hosts plus
subhalos) predicted by our subhalo model as a function of host  halo
mass, at redshift $z=0$.  The three panels show results for three mass
threshold values:  $\text{log}(\Mmin/\hMsun) = 11.4, 11.7$, and
$12.3$.  The four curves in each panel correspond to the four models
described in \S~\ref{model}: \emph{No Effects} considers no
gravitational effects on subhalos as they orbit inside  their host
halos (black dot-dashed curve); \emph{Fric. Only} considers only the
effects of dynamical friction (green  dashed curve);
\emph{Strip. Only} considers only the effects of mass loss (blue
dotted curve); \emph{Full} considers  both dynamical friction and mass
loss (solid red curve).  \emph{Right Panels}: The correlation function
of all halos  predicted by our subhalo model.  $\xi(r)$ is computed
from the halo model using the occupation statistics shown in the  left
panels.  The figure shows that dynamical effects (especially mass
loss) are needed in order to reduce the number of  subhalos
sufficiently and produce a power-law correlation function.  }
\label{fig:moment1_xi}
\end{center}
\end{figure*}

The left column of Figure~\ref{fig:moment1_xi} shows the mean
occupation number of host halos and subhalos as a function  of host
halo mass $\langle N \rangle_M$, at $z=0$.  The three panels give
results for halo samples defined by different  mass thresholds.  In
the interest of simplicity, we assume that all host halos and
surviving subhalos with masses  $M \ge \Mmin$ harbor an observable
galaxy.  This assignment is simpler than those supported by detailed
comparisons to  data, which typically assume that all host halos and
surviving subhalos with masses  $M \ge \Mmin$ (or some maximum
circular velocity) \emph{at the epoch of accretion} harbor an
observable galaxy.  We proceed in this manner because the subtleties
discussed in the aforementioned literature do not influence our
primary points and may serve to obscure them.  This is primarily
because any mass threshold chosen at the epoch of accretion will have
a second ``destruction'' threshold due to the finite resolution of a
given N-body simulation.  This can alter clustering measurements, and
since the aim of this paper is to present the qualitative trends
responsible for the low-redshift correlation function, we use the
simpler final mass approximation.  We have confirmed that using mass
at accretion with our model reproduces the same general results.  In a
forthcoming paper we consider more sophisticated models to compare
with data.  In the top, middle, and bottom panels we show samples with
$\log (\Mmin/\hMsun) = 11.4,\ 11.7$, and $12.3$, respectively.  These
particular mass thresholds result in average galaxy  number densities
(see Eq.~[\ref{eqn:ngal}]) equal to those in observed SDSS samples
with $r$-band luminosity thresholds  of $M_{r} < -18.5,\ -19.5,\text{
and } -20.5$ \citep{zehavi05a}.  The four curves in each panel
represent the four model  modes described in \S~\ref{model}, and we
calculate each curve from the mean of the 1000 model realizations.

First, the black dot-dashed curves represent the \emph{No Effects}
model.  As explained in \S~\ref{model}, this model  assumes that any
halo that merges into a larger host system (and becomes a subhalo of
that system) is thereafter  unaltered by dynamical effects in the host
halo environment.  Physically, this corresponds to the simple (and
observationally untenable) assumption that each subhalo above $\Mmin$
brings with it an observable galaxy upon merging  into the host and
that this galaxy is not destroyed or dimmed by dynamical evolution
within the host halo.  In effect,  each local peak of sufficient mass
in the smoothed density distribution forms a galaxy and the galaxy can
not be  destroyed.  Of course, we expect the mean halo occupation in
this model to be high for all host masses as compared to  the other
models.

Next, we turn to the curves depicting the individual effects of
dynamical friction (\emph{Fric. Only}, green dashed  curves) and mass
loss (\emph{Strip. Only}, blue dotted curves).  These dynamical
mechanisms can destroy subhalos, but  they cannot affect the host halo
or central galaxy.  This is why all the curves converge to the value
$\langle N\rangle_M =1$ at low host masses.  As a convenient
shorthand, we refer to any subhalo that fell into its host  system
with a mass $M_{\rm sub} \ge \Mmin$, but then merged with the central
host galaxy or lost sufficient mass to fall  below this threshold, as
\emph{destroyed}.  This does not mean that the subhalo has become
unbound, but merely that it  has either merged or no longer has a
bound mass above some minimum mass threshold.

Dynamical friction acting alone destroys subhalos by causing them to
sink to the centers of their hosts and ``merge''  with it.  This
mechanism alone causes a 20-35\% decrease in the mean number of
surviving satellites for all host masses  as compared to the \emph{No
Effects} model.  The fractional decrease in subhalos depends only
weakly on host mass, but  a comparison of the different mass threshold
panels shows a modest dependence on subhalo mass, with smaller mass
subhalos being depleted more.  These trends are counter-intuitive
because the dynamical friction force is an increasing  function of
$M_{\rm sub}/M_{\rm host}$, the mass ratio between the subhalo and its
host \citep{binney_tremaine08}.   One might expect the depletion of
subhalos to be larger for smaller host masses at fixed subhalo mass or
for larger  subhalo masses at fixed host mass.  Our results differ
from this expected behavior for two reasons.  First and foremost,
low-mass-ratio mergers tend to occur at higher redshifts than
high-mass-ratio mergers.  At higher redshifts, host halos  are
significantly smaller than they are at present, so high-redshift
mergers probe only the dense interiors of  contemporary host halos and
evolve approximately according to the subhalo-host mass ratio at the
redshift of the merger.   These early-merging subhalos also have a
longer period of time during which to evolve.  Second, our models
include  subhalos-of-subhalos.  As we move to larger host masses at
fixed subhalo mass or smaller subhalo masses at fixed host  mass, more
subhalos are subs-of-subs that have much higher mass ratios with their
immediate hosts.  These effects result  in the trends we see in
Figure~\ref{fig:moment1_xi}.

Mass loss is significantly more effective at ``erasing'' subhalos than
dynamical friction.  Mass loss processes can  effectively ``destroy''
subhalos because many lose sufficient mass to fall below the threshold
of a sample.  This  mechanism typically drives an 80-85\% decrease in
the number of objects above a given mass threshold compared to the
\emph{No Effects} model.  Again, the fractional decrease in subhalos
is nearly independent of host mass, but it shows  a slight dependence
on subhalo mass, with smaller mass subhalos being destroyed more
efficiently.

Finally, the \emph{Full} model (red, solid curves) includes the
effects of both subhalo mass loss and orbital decay by  dynamical
friction.  These processes do not simply sum together.  As a subhalo
sinks deeper into its host potential well  due to dynamical friction,
it experiences a stronger tidal field and is thus more efficiently
stripped of its mass.   Conversely, less massive subhalos are less
susceptible to orbital decay via dynamical friction.  A comparison of
the  \emph{Full} model to the \emph{Strip. Only} model shows that
including dynamical friction causes an additional  $\sim 15\%$
depletion of substructure.  Mass loss is by far the dominant cause of
subhalo destruction.  Overall,  Figure~\ref{fig:moment1_xi} shows that
dynamical effects reduce the number of subhalos by $\sim 90\%$
compared  to the number of distinct mergers that occur during the
formation of a host halo.

\subsection{Constructing the Correlation Function}
\label{xi_r}

We use the halo model outlined in \S~\ref{halomodel} to compute the
correlation function predicted by our subhalo model.   Specifically,
we use the \citet{jenkins01} mass function and we follow
\citet{tinker05} in using the \citet{smith03}  formula for the
non-linear matter power spectrum and the \citet{tinker05}
scale-dependent halo bias relative to the  non-linear power spectrum.
We derive HOD statistics from our subhalo models as exemplified by the
previous section,  and we compute the pair separation distributions by
assuming that the radial distributions of satellites follows an NFW
profile for simplicity.  In actuality, the subhalo distributions in
both our models and $N$-body simulations are slightly  shallower than
NFW (see Z05 for model and simulation results).  We adopt the NFW
profile for analytical convenience as  deviations from NFW are small
and only influence correlation functions notably on scales
significantly smaller than  $r \sim 100 \hkpc$ (e.g., Z05; also see
\citealt{watson_etal10} for a demonstration of this point regarding
satellite galaxies).

The right column of Figure~\ref{fig:moment1_xi} shows the host+subhalo
correlation functions computed in this manner  from the HODs predicted
by our subhalo model.  In the \emph{No Effects} case, where no
subhalos are destroyed, $\xir$  is very different from a power law,
having a one-halo term that is too large relative to the two-halo term
so that a  distinct feature is present in $\xir$ on scales $r \sim
2$~Mpc.  In fact, comparing this to Figure~\ref{fig:ximass},  we see
that it is very similar to the dark matter correlation function.  This
is perhaps not surprising because subhalos  in this model behave as
massive test particles that cannot be altered.  As dynamical effects
are included and  substructure is consequently depleted, $\xir$ drops
on all scales.  Recall that only subhalos (hosting satellite
galaxies) can be destroyed and the number of host galaxies remains
unaltered.  The fraction of all objects that are  satellites therefore
decreases.  As we discussed in \S~\ref{halomodel}, for a fixed
population of central galaxies,  the one-halo term drops in
approximate proportion to the number of satellite galaxies.  So as the
number of satellites  declines, so does the number of pairs within
halos relative to the total number of pairs and the one-halo term
declines.

Large-scale clustering is less sensitive to changes in the satellite
galaxy population.  The two-halo term drops because  subhalos tend to
populate more massive hosts (as in the left column of
Figure~\ref{fig:moment1_xi}), so the average host  halo mass of a
sample decreases as subhalos are depleted.  The large-scale clustering
strength of halos increases with  halo mass, so this depletion results
in weaker large-scale clustering.  The variability of the two-halo
term is  relatively mild because the halo bias is not a
rapidly-varying function of halo mass near $M \sim \Mmin$
\citep{tinker05}.

With enough depletion of substructure, the one- and two-halo terms
align and result in a nearly power-law shape.  This  is exactly what
happens in Figure~\ref{fig:moment1_xi}.  In our \emph{Full} subhalo
model the correlation function is  roughly a power law.  To obtain a
nearly power-law galaxy correlation function, it is necessary that a
majority of early  galaxies and proto-galaxies that merge to form a
massive system at low redshift be destroyed through either central
mergers or mass loss.  Our results suggest that mass loss is mainly
responsible for this depletion, while dynamical  friction and central
galaxy mergers play a comparably small, supporting role.
Incidentally, this picture implies that infalling satellite galaxies
lose  significant stellar mass so that they provide an important
source of the diffuse  intracluster light observed in galaxy groups
and clusters and this picture is consistent with observations
\citep{purcell_etal07,purcell_etal08}.  Comparing our correlation
function results for the three different mass  thresholds, we note
that $\xir$ is closer to a power law for lower-mass samples.  We
revisit this point in the next  section.

We now return to the mean occupation statistics shown in the left
panels of Figure~\ref{fig:moment1_xi}.  The so-called  ``plateau''
region of the HOD is the flat region at $\langle N\rangle_{M}=1$,
where host halos are more massive than  $\Mmin$, but not yet massive
enough to host subhalos above our mass threshold \footnote{Roughly
speaking, the most  massive subhalo within any host is a few percent
of the mass of the host halo (e.g., Z05).  This is the case with,  for
example, the Large Magellanic Cloud within the halo of the Milky Way
\citep{busha_etal10}.}.  As substructure is  depleted, the prominence
of this plateau increases.  The ``length'' of the plateau in the HOD
can be  expressed as the ratio between the mass of a halo that hosts a
single satellite on average, $\Mone$, to the minimum  mass required to
host a central galaxy, $\Mmin$.  \citet{zehavi05a} fit an HOD model to
the measured correlation  function of SDSS galaxies and found that a
ratio $\Mone/\Mmin \sim 23$ is consistent with clustering data, nearly
independent of galaxy luminosity.  In other words, a consistent
picture is one in which the entire HOD shifts to  higher masses in a
self-similar manner, with $\Mone/\Mmin$ fixed, in order to accommodate
higher-luminosity samples.   Remarkably, \citet{kravtsov04a} studied
this for subhalos in a high-resolution $N$-body simulation and found
that  $\Mone/\Mmin \sim 20$, regardless of $\Mmin$ as well.
Meanwhile, \citet{tinker05} fit a slightly more complex HOD  model to
the SDSS data and found that $\Mone/\Mmin \sim 25$ for galaxy samples
with luminosities less than $\Lstar$,  but decreases to $\Mone/\Mmin
\lesssim 5$ to accommodate the highest-luminosity samples (absolute
$r$-band magnitudes  $M_r\leq -21$).  The new analysis by
\citet{zehavi10} also finds this trend with $\Mone/\Mmin \sim 17$ for
$M_r\geq -20.5$ and much lower values for higher luminosity galaxies.
For the purpose of comparison, the \emph{Full}  subhalo model shown in
Figure~\ref{fig:moment1_xi} predicts $\Mone/\Mmin \sim 40$ for the
low-mass samples of  $\log(\Mmin/\hMsun)=11.4$ and
$\log(\Mmin/\hMsun)=11.7$, and $\Mone/\Mmin \sim 30$ for the
higher-mass sample of  $\log(\Mmin/\hMsun)=12.3$.

These results suggest that getting the length of the HOD plateau right
may be a key ingredient needed  to establish a power-law correlation
function and this has been part of the interpretation in the
literature.  The importance of $\Mone/\Mmin$ stems from the fact that
most one-halo pairs reside  in halos with average satellite numbers
$\langle N_{\rm s} \rangle_M \lesssim 1$, so modeling the HOD  at
relatively low satellite occupation numbers is critical
\citep[see][]{conroy06}.  We investigate this further in
\S~\ref{make_powerlaw}.

\section{MASS AND REDSHIFT DEPENDENCE OF THE CORRELATION FUNCTION}
\label{dependence}

\subsection{Dependence on Mass}
\label{mass_dependence}

\begin{figure}
\begin{center}
\includegraphics[width=.5\textwidth]{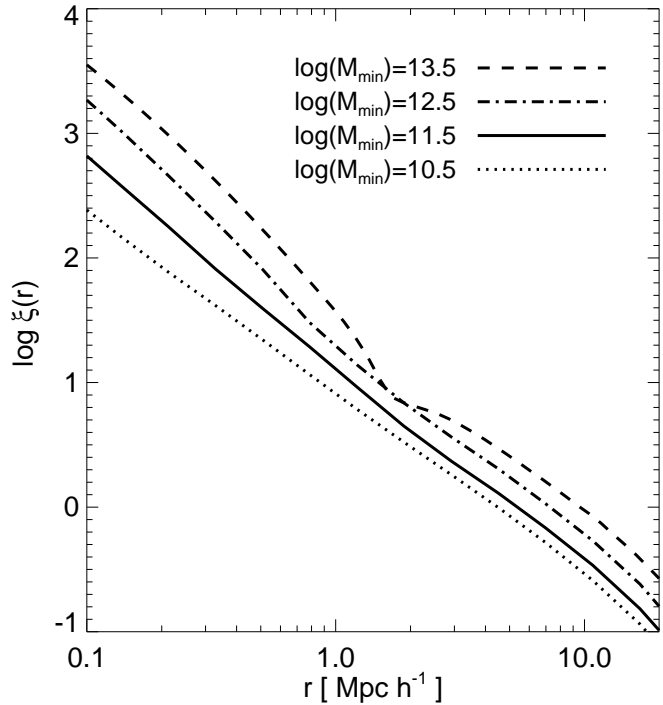}
\caption{ Correlation function of all halos (hosts plus subhalos)
predicted by our subhalo model at redshift $z=0$.  The four  curves
show $\xir$ for four mass threshold samples and the threshold values
$\Mmin$ (in units of $\hMsun$) are listed  in the panel.  The figure
shows that $\xir$ breaks more and more from a power law for higher
mass halo samples, which  correspond to higher luminosity galaxy
samples.  }
\label{fig:xi_5Mmin_cuts}
\end{center}
\end{figure}

While dynamical processes act in a manner to deplete substructure and
push $\xir$ toward a power law at all mass  thresholds, it is evident
that deviations from a power law are stronger with increasing host
mass.   Figure~\ref{fig:xi_5Mmin_cuts} shows the correlation functions
predicted by our \emph{Full} subhalo model for four  different mass
thresholds, ranging from $\log(\Mmin/\hMsun)=13.5$, corresponding to
bright galaxies such as Luminous  Red Galaxies (LRGs), down to
$\log(\Mmin/\hMsun)=10.5$, corresponding to dwarf galaxies.  While the
correlation function  of the ``dwarf'' sample is a near power law,
that of the ``LRG'' sample exhibits strong departures from power-law
behavior.  This trend has been detected with SDSS galaxies by
\citet{zehavi05a} who found evidence that a power-law  model provides
a better fit to low-luminosity galaxies than high-luminosity galaxies.
Halo and subhalo clustering  exhibits the same trend.  More massive
halos contain slightly more of their bound masses in substructure
relative to less  massive halos, but this is a comparably small effect
(Z05) and drives only $\sim 30\%$ of the mass-dependence of  the
one-halo term in Figure~\ref{fig:xi_5Mmin_cuts}.  At {\em fixed
redshift}, the departure from a power-law at  high mass (high
luminosity) is caused by the relative rareness of high-mass host halos
(see \S~\ref{halomodel}  for interpretive discussion).

\subsection{Dependence on Redshift}
\label{redshift_dependence}

Substructure abundances vary with time.  Infall of new subhalos acts
as a ``source'' of halo substructure.  The rate of  mergers of halos
into larger systems is a function of redshift that typically peaks at
redshifts $z \sim 1-3$ in the  halo mass range of interest and
declines thereafter (Z05; \citealt{zentner07}).  Once a subhalo merges
into a larger  host halo, dynamical friction shrinks its orbit and the
subhalo loses mass.  Given enough time, the subhalo will  eventually
lose enough mass to fall below $\Mmin$ or merge with the central
galaxy and lose its identity.  The balance  between the halo merger
rate and the rates of destructive processes (which occur on a halo
dynamical time) determine  the redshift dependence of halo
substructure.

\begin{figure*}
\begin{center}
\includegraphics[width=\textwidth,angle=0]{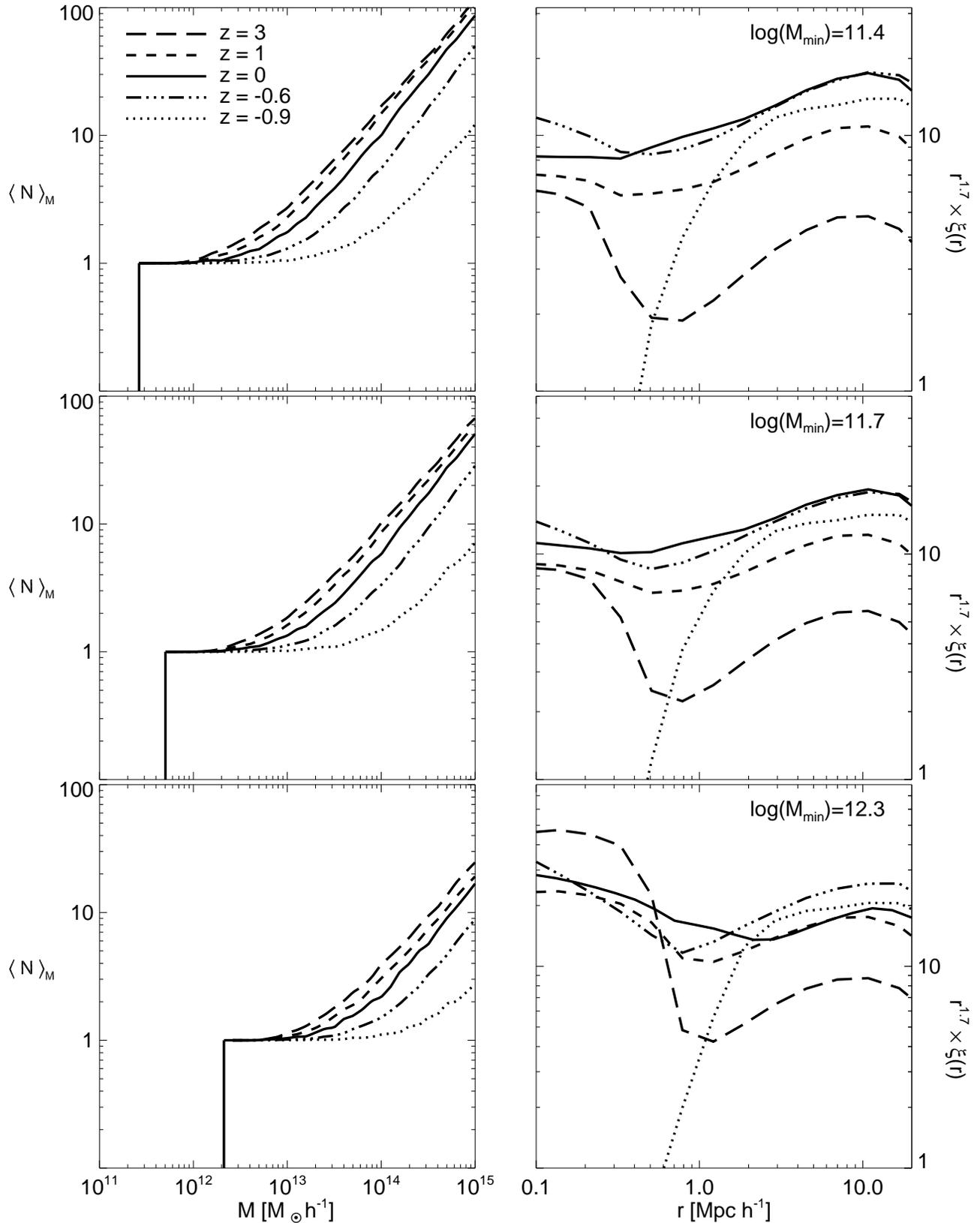}
\caption{ \emph{Left panels}: Mean number of all halos (hosts plus
subhalos) predicted by our \emph{Full} subhalo model as a  function of
host halo mass, at five different redshifts.  The three panels show
results for three mass threshold values:  $\log (\Mmin/\hMsun) =
11.4,\, 11.7$, and $12.3$.  The five curves in each panel correspond
to the redshifts  $z=3, 1, 0, -0.6, -0.9$ (negative redshifts
correspond to future epochs).  \emph{Right panels}: Correlation
functions  corresponding to the halo samples shown in the left panels.
In each case, $\xir$ has been scaled by a power law in  order to
clearly show departures from a power-law shape.  The figure shows that
the number of subhalos steadily  decreases from high to low redshift,
causing the correlation function to evolve from not being a power law
at high  redshift, towards having a nearly power-law shape at the
present epoch, and once again deviating from a power law at  future
epochs.  }
\label{fig:xi_evolution}
\end{center}
\end{figure*}

Figure~\ref{fig:xi_evolution} shows the redshift evolution of the mean
halo occupation number and resulting correlation  functions.  The
layout of Figure~\ref{fig:xi_evolution} is similar to that of
Figure~\ref{fig:moment1_xi}, with  $\langle N\rangle_M$ shown in the
left panels, $\xir$ shown in the right panels.  However, in
Figure~\ref{fig:xi_evolution} all results are for the \emph{Full}
subhalo model, and the various lines denote  quantities evaluated at
different redshifts, $z=3,\ 1,\ 0,\ -0.6,\ -0.9$ (where negative
redshifts correspond to  \emph{future} epochs).  Moreover, in each
panel the correlation functions are scaled by a power law to better
highlight  departures from a power-law shape.

The left-hand panels of Figure~\ref{fig:xi_evolution} show that the
average number of subhalos within hosts of a given  mass starts out
high at early times and begins to decrease after $z=3$, as merger
rates decline.  By the present epoch  ($z=0$), the number of subhalos
has dropped by $\sim 25-30\%$ relative to what it was at $z = 3$.  One
Hubble time into  the future ($z=-0.6$), the abundance of substructure
has dropped by $\sim 60\%$.  This is because the rate of merging as  a
source for new subhalos declines rapidly. This decrease in the merger
rate is dictated in large part  by the quenching of structure growth
by the cosmological constant \citep{carroll92}, but also because most
halos of  interest are below the typical collapsing mass, which
approaches $\Mstar \simeq 10^{14}\, h^{-1}\mathrm{M}_{\odot}$ in  the
future \citep{zentner07}.  Meanwhile, destructive processes continue
to operate on orbiting halo substructure for  several additional
dynamical times.  Three Hubble times into the future ($z=-0.9$) the
average halo occupation has  dropped by $\sim 90\%$.  As with our
previous results, the fractional decrease in subhalo abundance appears
to be  roughly independent of host mass, meaning that the slope of the
HOD in the high-$\langle N \rangle_M$ limit is not  significantly
altered by evolution.  The amplitude of $\langle N\rangle_M$ declines
considerably, resulting in increasing  $\Mone/\Mmin$, or a
``lengthening'' of the HOD plateau with time.  This behavior is
strikingly similar to that seen in  Figure~\ref{fig:moment1_xi} in the
sense that turning on dynamical effects at a fixed redshift has a
qualitatively  similar impact as evolving forward in time, and the
effects on the correlation function are similar.

Turning to the right panels, $\xir$ is shown at each redshift scaled
by an $r^{-1.7}$ power law in order to emphasize  features in the
correlation function.  Starting at $z=3$ (long dashed curves), $\xir$
is very far from a power-law, with  a slope that is much steeper on
small scales.  At $z=1$ (short dashed curves) the break from a power
law is less  pronounced, but it is still significant.  These results
are qualitatively consistent with clustering measurements at  high
redshifts \citep{coil06,ouchi05,lee06}.  At $z=0$ (solid curves), the
correlation function is approximately a power  law, though there is
still a mild, discernible feature at the transition scale between the
one- and two-halo terms.   In the future, $\xir$ once again breaks
from a power law.  At $z=-0.6$ (dot-dashed curves), departures from a
power-law  shape are about as strong as they were at $z=1$.  Three
Hubble times into the future, at $z=-0.9$ (dotted curves), the
departures from a power law are significant and represent a dramatic
reduction in the relative contribution of the  one-halo term.

\begin{figure}
\begin{center}
\includegraphics[width=.5\textwidth,angle=0]{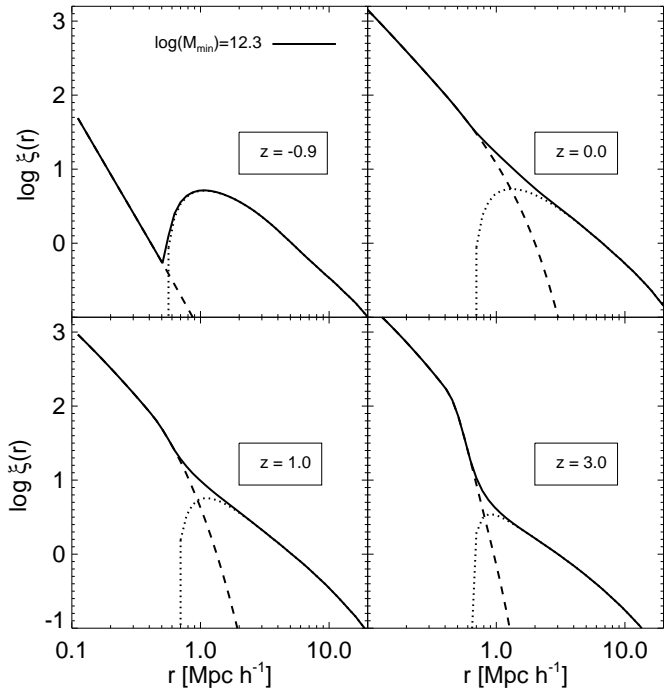}
\caption{ The correlation function of all halos (hosts plus subhalos)
predicted by our \emph{Full} subhalo model as a function of  redshift,
for a single mass threshold sample $\text{log}(\Mmin/\hMsun)=12.3$.
Each panel shows $\xir$ for a different  redshift (solid curve), as
well as the one-halo (dashed curve) and two-halo (dotted curve) terms.
The figure shows that  the one-halo term evolves strongly with
redshift and only at $z=0$ strikes the right balance with the two-halo
term to  result in a power law.   }
\label{fig:xi_r_contribution}
\end{center}
\end{figure}

Figure~\ref{fig:xi_r_contribution} focuses on the $\log(\Mmin/\hMsun)
= 12.3$ threshold sample and shows the correlation  function at four
different redshifts, while also showing the one-halo and two-halo
terms explicitly.   Figure~\ref{fig:xi_r_contribution} clearly
demonstrates how a delicate balance is needed between the two terms in
order  for $\xir$ to achieve a power-law shape.  The two-halo term
exhibits modest variations from panel to panel, with a range  of about
a factor of $\sim 3$.  The decreased large-scale clustering at $z
\gtrsim 0$ is due to  the linear growth of perturbations with time,
but this is always kept modest because the increasing  bias of halos
of fixed mass with redshift \citep[see][]{zentner07} compensates for
large-scale structure  growth.  At $z < 0$, the slight decrease in
two-halo clustering is due to the decay of halo bias once  halo growth
slows \citep{fry96}.

The variation in the one-halo term is significantly larger, as our
earlier discussions  suggest, and changes by a factor of $\sim 45-150$
(depending on scale), equivalent to $\sim 15-50$ times  the variation
in the two-halo term.  At high redshift, the relative rareness of
host halos and the large amount of substructure cause $\xir$ to be
boosted significantly in the one-halo regime  as shown in the $z=3$
panel of Figure~\ref{fig:xi_r_contribution}.  At $z = 0$, just  the
right amount of substructure has been depleted to strike a near
balance between the one-halo and two-halo  contributions.  In the
future, the continual destruction of subhalos suppresses the one-halo
term, driving $\xir$ away  from a power law again.  By $z=-0.9$, the
depression in small-scale clustering is striking.

Some of the evolution of $\xir$ on small-scales comes from the fact
that halos large enough to  host luminous galaxies become increasingly
rare as redshift increases.   The characteristic collapsing mass is a
rapidly decreasing function of  redshift and is only $\Mstar \approx
10^9\, h^{-1}\mathrm{M}_{\odot}$ at $z = 3$.  In the relevant regime,
the strength  of the one-halo term grows in approximate proportion to
the number of satellite galaxies and in inverse proportion to  the
number of host halos of appropriate size (see \S~\ref{halomodel}), so
the relative paucity of host halos at high  redshift also drives
strong one-halo clustering because Fig.~\ref{fig:xi_r_contribution}
describes samples of fixed  absolute mass threshold.  However, it is
subhalo abundance that has the larger influence on the redshift
dependence of  clustering.  We have computed the correlations of
Figure~\ref{fig:xi_r_contribution} using samples in which $\Mmin$
varies with redshift so as to maintain a {\em constant number density}
of halos.  These samples are less subject to the  gross evolution of
the halo mass function.  We find all of the same qualitative results
for this case, though the  two-halo term varies by a factor of $\sim
4$, while the variation in the one-halo term is limited to a factor of
$\sim 12-80$ (again, depending on scale), resulting in a variation in
the one-halo term that is $\sim 3-20$ times  larger than that of the
two-halo term.  Moreover, we have re-computed correlation functions
using a combination of the predicted low-redshift HODs alongside the
high-redshift mass functions in order to isolate the  contribution due
to the mass function and HOD evolution.  The majority the redshift
dependence of $\xir$ on small  scales is due to the evolution of
subhalo abundance.  To maintain a  power-law correlation function at
high-redshift would require fewer subhalos per host than at $z=0$ in
order to  compensate for the relative rareness of host halos at
high-redshift.  In fact, hosts at high redshift have a {\em larger}
number of subhalos of any given mass so these effects {\em reinforce
one another}, leading to a strong deviation from a  power-law $\xir$
at high redshift.

We have already described the reasons that the one-halo and two-halo
terms behave so differently under changes in the  HOD.  To reiterate,
on large scales, $\xir$ is essentially a weighted average of the
clustering of host halos, where  $\langle N \rangle_M$ provides the
weighting (see the integral in Eqs.~[\ref{eqn:xi_2h}]
and~[\ref{eqn:xi_2h_lim}],  note that $\tilde{\lambda}(k,M) \approx 1$
for $k < 1/R_{\mathrm{vir}}$).  The possible variability in $\xir$ on
large  scales is limited because it is always bound by the limited
variation in the clustering of host halos.  As we discussed  in
\S~\ref{halomodel}, the difference in the large-scale bias of the
largest relative to the smallest halos is at most  a factor of $\sim
3$ (e.g., \citealt{tinker05}).  Significant variations in large-scale
clustering  require dramatic variations in the HOD at high mass, which
are not expected on theoretical grounds and are not mandated  by data.
However, on scales smaller than the size of individual host halos,
$\xir$ can vary dramatically depending on  the HOD.  For example, in
the extreme case of only one object per host halo, there will be zero
pairs within halos and  the one-halo term will vanish.  For large
numbers of satellites, the one-halo term will be significantly larger
than a  power-law extrapolation of the two-halo term to small scales.

The sensitivity of the one-halo term to the HOD, coupled with the
relative insensitivity of the two-halo term, means  that achieving a
power-law correlation function requires fine-tuning in the number of
satellite galaxies per halo.  The  satellite galaxy abundance
naturally evolves with redshift, so $\xir$ can only be a power law
during those epochs when  substructure has evolved to join the
one-halo term to the two-halo term.  Of course, it may be possible for
features in  the host halo mass function or bias relations to conspire
to compensate for substructure evolution, but such features  would
somehow need to be coordinated with low-redshift structure growth.  A
different way to state this is that the halo  mass function and halo
bias depend on the statistics of the linear density field, and do not
``know'' about the  non-linear galaxy formation and gravitational
processes that occur within halos.   It would be quite strange if
their evolution were somehow connected with the evolution of
satellite galaxies in virialized hosts.  It seems to be a coincidence
that the epoch of  near power-law clustering of typical galaxies lies
near $z=0$.

\subsection{The Balance Between Accretion and Destruction}
\label{accr_destr}

\begin{figure}\label{merge_destr}
\begin{center}
\includegraphics[width=.5\textwidth,height=.4\textwidth,angle=0]{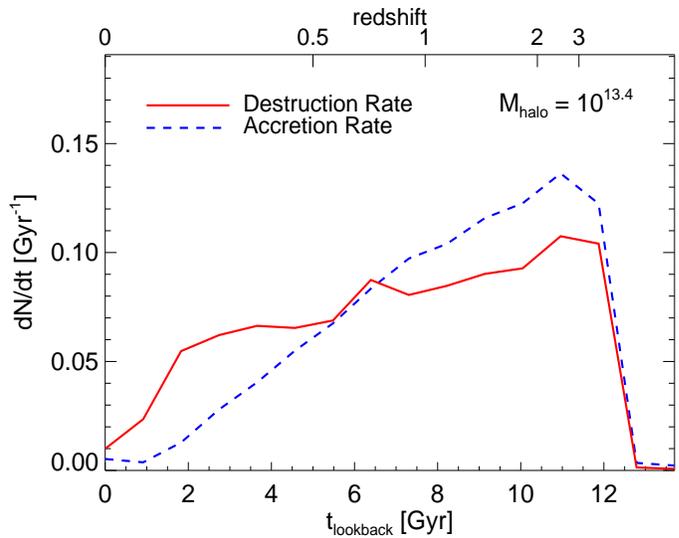}
\caption{ The accretion versus destruction rate of subhalos over
cosmic time, as predicted by our full subhalo model.  The  accretion
rate shown is the number of subhalos per Gyr that merge into a host
halo of mass $\text{log}(M/\hMsun)=13.4$.   The destruction rate is
the number of these same subhalos per Gyr that are destroyed (i.e.,
their mass drops below some  threshold value).  The two rates
equalized when the Universe was $\sim 6$Gyr old (at $z\sim 1$).
Before $z=1$, the net  number of subhalos increased with time, whereas
at later times the net number decreased with time.  The figure shows
how  the balance between accretion and destruction changes with
redshift, which explains why the correlation function can  only be a
power law at a single epoch.  }
\label{fig:accr_destr}
\end{center}
\end{figure}

We have just seen how the depletion of substructure over time leads to
evolution in the correlation function such that  it becomes a power
law at the present epoch.  However, what drives substructure
depletion?  We expect that most subhalos  will lose significant
amounts of mass or merge with the central galaxy given sufficient
time.  However, this will be  compensated to some degree by the infall
of new subhalos.  If the rate at which satellites are accreted is
greater than  the rate at which they are destroyed, then the net
amount of substructure will grow with time.  The evolution in the
number of subhalos (and hence the correlation function) depends on the
balance between accretion and destruction.  Z05  give a related
discussion of accretion and destruction in their \S~4.4 and the
perspective we adopt here complements Z05.

In Figure~\ref{fig:accr_destr} we illustrate the competition between
accretion and destruction in host halos of mass  $M = 10^{13.4}\
h^{-1}\mathrm{M}_{\odot}$.  To measure the accretion rate (dashed
curve), we count all subhalos with masses  greater than
$10^{11}\hMsun$ that accrete onto these hosts in finite time
intervals.  For the  destruction rate (solid curve), we count the
number of these same subhalos that drop below $10^{11}\hMsun$ during
the  time intervals.  The accretion rate minus the destruction rate
will then give us the net rate of change in the number  of subhalos
per unit time.

Figure~\ref{fig:accr_destr} shows that the accretion rate quickly grew
and reached a peak at $z\sim 2-3$.  Since this  peak, the accretion
rate has been steadily declining and is close to zero at the present
epoch.  The decline in merger  rates is partly due to the shape of the
power spectrum \citep[see][]{LaceyCole93,somerville99,zentner07}, but
the  driving force for the recent fast decline in the merger rate of
halos is the reduced rate of structure growth caused by  accelerated
cosmic expansion.  The destruction rate also peaked at $z \sim 2-3$
and lags the accretion rate because most  destruction happens over a
period of several dynamical times.  Figure~\ref{fig:accr_destr}
clearly shows that the  accretion rate has been dropping faster than
the destruction rate since their peaks, with accretion and destruction
roughly balancing just below $z \sim 1$ \citep[see also][]{stewart09}.
This means that the number of subhalos in  hosts that grow to a mass
of $M=10^{13.4}\ h^{-1}\mathrm{M}_{\odot}$ by $z=0$ increased until $z
\approx 1$ and has been  declining ever since, despite the fact that
the virial masses of these halos have been growing.  The general trend
toward reduced substructure at low redshift explains the behavior
exhibited in Figure~\ref{fig:xi_evolution}.  The  correlation function
is close to a power law at the present epoch because the balance
between accretion and  destruction over time has led to the requisite
abundance of substructure today.

\section{Achieving a Power-Law Correlation Function}
\label{make_powerlaw}

We now step back from making predictions using our specific subhalo
model and undertake a general exploration of the  properties of the
HOD that yield nearly power-law correlation functions at different
masses and redshifts.  The HOD  characterizes the number and spatial
distribution of galaxies within dark matter halos.  It is typically
specified  with a handful of parameters that are constrained using
galaxy clustering measurements
\citep[e.g.,][]{magliocchetti03,zehavi05a,tinker05,zheng07}.  We
choose an HOD model that is motivated by  theoretical predictions from
hydrodynamic simulations, semi-analytic models, and high-resolution
N-body simulations  \citep{berlind03,kravtsov04a,zheng05}.  According
to this model, halos above some threshold mass contain a single
``central'' galaxy plus a number of ``satellite'' galaxies.  The
number of satellites in any given halo is drawn from  a Poisson
distribution whose mean is a power-law function of host halo mass.
The central galaxy is placed at the  center of the host halo, while
the satellites are spatially distributed according to an NFW density
profile.   Specifically, we adopt an HOD parametrization that is
similar to the one used by \citet{tinker05}.  This is a  simple, yet
powerful model in which the number of central galaxies is modeled as a
step function,
\begin{equation}\label{eqn:MeanN}
N_\mathrm{cen} = \left\{
	\begin{array}{ll}
		 1 &\mbox{if }  M \geq M_\mathrm{min}\\ 0 &\mbox{if }
		 M < M_\mathrm{min}
	\end{array}
\right. ,
\end{equation}
while the mean number of satellites follows a power-law with an
exponential cutoff at low mass,
\begin{equation}\label{eqn:MeanNsat}
\langle N_\mathrm{sat} \rangle_M =
\Bigg(\frac{M}{\Mone}\Bigg)^{\alpha}\text{exp}\Bigg(\frac{-\Mzero}{M}\Bigg).
\end{equation} 
The parameters in the model are as follows.
\begin{enumerate}
\item $\Mmin$ is the minimum host halo mass to contain a central
galaxy.
\item $\Mzero$ is the host halo mass below which satellite galaxies
are exponentially supressed.
\item $\Mone$ is the host halo mass to contain, on average, one
satellite galaxy.
\item $\alpha$ is the index of the power-law relation between the mean
number of satellite galaxies and halo  mass.
\end{enumerate}

Previous studies have shown that the power-law index $\alpha \approx
1$ for subhalos and simulated galaxies
\citep{kravtsov04a,zheng05,zentner05}, as well as observed galaxies
dimmer than $\Lstar$ \citep{zehavi05a}, leading  \citet{tinker05} to
set $\alpha = 1$ throughout their analysis.  However, we allow
$\alpha$ to vary because the  correlation function is sensitive to it
and, although it may be near unity when modeling observed data, it may
need  to deviate from unity to yield a power-law correlation function
at high redshifts.  On the other hand, $\xir$ is not  sensitive to
$\Mzero$, consequently we fix its value by adopting the
\citet{conroy06} $\Mzero - \Mone$ relation,
\begin{equation}\label{eqn:Mzero}
\mathrm{log}(\Mzero /\hMsun) = 0.76 \ \mathrm{log}(\Mone /\hMsun) +
2.3.
\end{equation} 
The result is an HOD model with only three free parameters: $\Mmin$,
$\Mone$, and  $\alpha$.

The two-halo term of $\xir$ depends on the mean occupation $\langle N
\rangle_M = \langle \Ncen+\Nsat \rangle_{M}$,  which is equal to
$1+\langle \Nsat \rangle_{M}$ for $M>\Mmin$.  The one-halo term also
requires the second moment of  the occupation distribution $\langle
\Nsat(\Nsat - 1)\rangle_{M}$, so characterizing the mean occupation is
not  sufficient.  We assume that the number of satellites follows a
Poisson distribution, for which  $\langle \Nsat(\Nsat - 1)\rangle_{M}
\equiv\langle \Nsat \rangle^2_{M}$.  Our \emph{Full} model deviates
mildly from  a pure Poisson distribution \citep[see Fig.~7 of Z05, and
recent simulations of][that find similar deviations from a  Poisson
distribution]{bk10}, but the effect of this deviation on $\xir$ is
minor (Fig.~16 of Z05).  We also note that  there are any number of
possible parametrizations for $\langle \Nsat\rangle_M$ to choose from
besides the one adopted  here.  We have found that mildly different
parametrizations that exhibit the same basic features and are
consistent  with contemporary data (e.g., the one used by
\citealt{zehavi05a}) yield similar conclusions.

\begin{figure*}
\begin{center}
\includegraphics[width=\textwidth,angle=0]{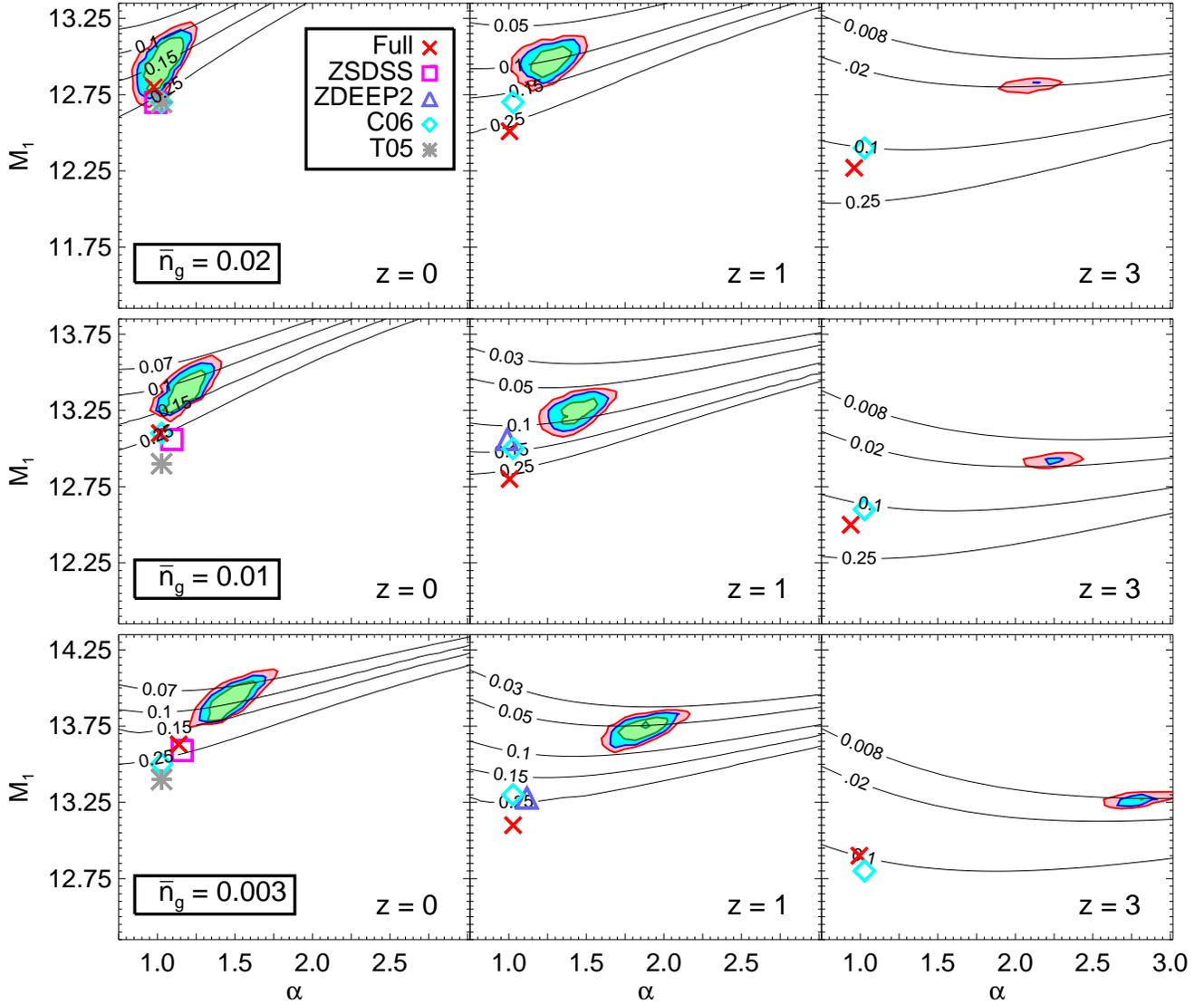}
\caption{ Exploration of the HOD parameter space that yields a
power-law $\xir$, as a function of redshift and sample number
density.  Each column of panels shows results for a different redshift
($z=0, 1, 3$).  Each row of panels shows results  for a different
sample number density ($\ngal=0.02, 0.01, 0.003 \hmpcVol$).  We adopt
the four-parameter HOD model  shown in
equations~\ref{eqn:MeanN},~\ref{eqn:MeanNsat}, and~\ref{eqn:Mzero}.
Each panel shows the parameter space  probed by $\alpha$, the slope of
the mean occupation number of satellites, and $\Mone$, the halo mass
that contains  on average one satellite galaxy.  For each pair of
$\alpha$ and $\Mone$ values, we find the value of $\Mmin$ that  yields
the desired galaxy number density.  We then use the halo model to
compute $\xir$ for that set of HOD parameters.   We do this on a
$50\times 50$ grid of $\alpha - \Mone$ parameter combinations.  We fit
each correlation function to a  power law, and the shaded contours
represent the $68.3\%, 95\%$ and $99.6\%$ power-law likelihood (green,
blue, red  contours).  Also shown are contours of constant satellite
fraction (solid black curves). The red X in each panel shows  the HOD
parameters predicted by our \emph{Full} subhalo model.  For
comparison, we also show results from HOD modeling  of real galaxy
samples from the SDSS at $z=0$ (\citealt{zheng07}, T05 - magenta boxes
and grey asterisks); and DEEP2 at  $z=1$ (\citealt{zheng07} - purple
triangles).  Finally, we show the simulation results of
\citet{conroy06} that are  designed to model SDSS, DEEP2, and
Lyman-break galaxies at $z=0$, 1, and 3, respectively (cyan diamonds).}
\label{fig:fixedNumDens}
\end{center}
\end{figure*}

We consider the HOD parameter space that yields a power-law
correlation function for three galaxy samples of fixed  number density
$\ngal$, at three different redshifts $z=0, 1, 3$.  Fixing number
density is a way to compare similar  samples at different redshifts
because the high-redshift sample is more likely to represent the
progenitors of the  low-redshift sample than it would in the case of
mass threshold samples.  We choose number densities equal to
$\ngal=0.02$, $0.01$, and $0.003\ \hmpcVol$, which correspond to three
$z \simeq 0$, volume-limited,  $r$-band threshold samples in the SDSS:
$M_{r} < -18.5,\ -19.5$, and $-20.5$ \citep{zehavi05a}.

For a given number density and redshift combination (e.g., $\ngal\,
=\, 0.02\ \hmpcVol$ at $z=1$), we create a  $50 \times 50$ grid of
$\Mone -\alpha$ parameter combinations.  For each pair of $\Mone$ and
$\alpha$ on this grid, we  use Equation~(\ref{eqn:ngal}) to find the
value of $\Mmin$ that is needed to enforce the desired number density.
In this manner, the 2,500 HOD models on the grid represent galaxy
samples with the same number densities,  but different HODs.  We then
compute the first and second moments of the mean galaxy occupation
using equations~(\ref{eqn:MeanN}) and~(\ref{eqn:MeanNsat}), and use
the halo model described in  \S~\ref{halomodel} to construct $\xir$.
We assign $10\%$ errors on all scales to $\xir$, as such errors are
roughly consistent with  jackknife re-sampling errors in current
clustering measurements \citep{zehavi05a}, and we  fit a power-law
function to all 2,500 correlation functions.   We perform our fits
using a Markov Chain Monte Carlo (MCMC) analysis in which we vary the
slope and correlation  length of the fitted power-law.  We then find
the minimum $\chi^{2}$ value for a power-law fit to $\xir$ for any
given  $\Mone-\alpha$ combination.  This allows us to approximate the
HOD parameter space in which  $\xir$ is consistent with a power law at
a level similar to contemporary observations.   For two free
parameters, the $68.3\%$ ($1\sigma$), $95\%$ ($2\sigma$), and $99.6\%$
($3\sigma$)  likelihood regions correspond to values of reduced
$\chi^{2} \leq 1.15,\ 1.61$, and $2.06$,  respectively.

Figure~\ref{fig:fixedNumDens} shows the contours generated from the
aforementioned procedure.  Each row in the figure  represents a
different $\ngal$ value and each column corresponds to a different
redshift.  The ``satellite fraction''  ($\fsat$, the fraction of all
galaxies that are satellites, see Eq.~[\ref{eqn:fsat}]) is relevant to
the shape of the  galaxy two-point correlation function.  Therefore,
over-plotted in each panel are curves of constant $\fsat$ (the
labeled, solid, black curves).  To compare these results with
measurements from observed galaxies, in each panel we  also show
best-fit $\Mone$ and $\alpha$ values from published halo model fits to
measurements of $\xir$ using galaxy  samples with the same number
densities at the same redshifts.  Squares and triangles represent the
best-fit parameter  values from \citet{zheng07} who fit SDSS ($z=0$)
and DEEP2 ($z=1$) data (ZSDSS, ZDEEP2) and asterisks represent the
best-fit \citet{tinker05} values for SDSS data (T05).  Diamonds
represent the \citet{conroy06} values  for SDSS, DEEP2, and the $z=3$
Subaru data of Lyman-break galaxies (C06).

The best-fit parameter combinations should be regarded as best-fit
``regions'', because there are errors associated  with the derived
parameters (e.g., the \citet{zheng07} SDSS $\alpha$ and $\Mone$ errors
at each luminosity are of order  $10\%$). We note that
\citet{tinker05} considered several possible values of $\sigma_{8}$,
but we show their results  for $\sigma _{8}=0.9$ to be consistent with
the cosmological model used in the other studies.  Finally, in each
panel  we show the HOD parameters predicted by our {\em Full} subhalo
model (marked by an ``X'') for samples with mass  thresholds that
yield the desired number density.  The {\em Full} model gives $\langle
\Nsat \rangle_M$ and we fit this  with Equation~(\ref{eqn:MeanNsat})
to obtain best-fit values of $\Mone$ and $\alpha$.

Several interesting conclusions can be drawn from this figure.
\begin{enumerate}\label{fixedngal_summary}
\item{The region of HOD parameter space that yields a power-law $\xir$
drifts to lower values of both $\Mone$ and  $\alpha$ with increasing
number density.  These trends increase the satellite fraction as
number density increases  to compensate for the relative reduction in
the one-halo term compared to the two-halo term induced by moving to a
lower-mass, more abundant halo sample.}
\item{The values of $\alpha$ that result in the best power laws drift
higher with increasing redshift in an effort to  boost the two-halo
term by placing galaxies in massive, highly-biased halos.   In
general, it is difficult to arrange a power law at $z \ge 3$ for these
three number densities. }
\item{As might be expected from our previous discussions, there is a
relatively narrow range of $\fsat$ for the best-fit  power-law space
at each redshift.  At $z=0$, the space that is consistent with a
power-law with $10\%$ errors on the data  lie near $\fsat \sim
0.1-0.15$.  The direct fits to observational data lie near $\fsat=0.2
- 0.3$.   At $z=1$, the power-law region is shifted to $\fsat \sim
0.05-0.1$, while at $z=3$ the power-law region is even  lower, $\fsat
\sim 0.01-0.02$.  Note that the power-law regions are not precisely
aligned along constant-$\fsat$  contours, particularly at low redshift
and low number density, indicating that other factors, such as host
halo  abundances and the physical sizes of host halos, contribute to
the power-law nature of $\xir$.  However, at high redshift and low
number density, the power-law regions become more nearly  co-linear
with contours of constant $\fsat$ over a range of $\alpha$ values.}
\item{The SDSS ($z \sim 0$) best-fit points lie near the power-law
contours, but not within these likelihood regions.   This is not
surprising as the SDSS measurement is more precise over a wide range
of scales than the $\sim 10\%$ errors  we have assumed and the
observed $\xir$ is now known to exhibit very small, but
statistically-significant deviations  from a power law
\citep{zehavi04a}.   }
\item{As predicted from Figure~\ref{fig:xi_5Mmin_cuts}, the fits to
observational data lie further from the  power-law regions as we move
to lower $\ngal$ (higher luminosity) samples.  At fixed redshift, this
is driven largely  because the host halos of these galaxies become
increasingly rare.   However, it is worth noting that the growth of
the one-halo term  with increasing $\Mmin$ is reinforced by an
increase in satellite abundance  at fixed scaled mass $\Msub/\Mhost$
as $\Mhost$ increases, accounting for  $\sim 30\%$ of the rise.  This
increase satellite abundance with $\Mhost$ arises because  more
massive host halos assemble more recently, leaving less time for the
evolution of  substructure and less satellite destruction (Z05).  The
relative time available  for satellite evolution is an important part
of  determining the power-law nature of the correlation function.  }
\item{The fits to observational data lie near the power-law regions at
$z=0$, but grow more distinctly separated with  increasing redshift.
This evolution is driven by satellite fractions at high-$z$ that are
too large to be consistent  with power-law clustering.  This supports
our basic picture that satellite destruction over cosmic time is
needed to  achieve a power law $\xir$, and that the observed
low-luminosity, low-redshift $\xir$ is a coincidence.  }
\item{The HOD values predicted by our \emph{Full} subhalo model are
similar to all of the observed data fits at all  redshifts.  This is a
remarkable result considering our model treats only subhalos and not
galaxies explicitly.   We explore more complicated associations of
galaxies and subhalos in a follow-up study.  }
\item{Our subhalos, as well as all observational data, reveal values
of $\alpha \simeq 1$  for all redshifts, in accord with previous
theoretical results \citep{kravtsov04a,zheng05,zentner05,conroy06}.
Moreover, at each redshift, they have fixed satellite fractions,
independent of $\ngal$.  At $z=0$, $z=1$, and $z=3$,  our model and
the observational data cluster near $\fsat \approx 0.25$, $\fsat
\approx 0.2$, and $\fsat \approx 0.1$,  respectively.  We note that
the lower satellite fractions at high redshift are {\em not} due to
HOD evolution.   Figure~\ref{fig:xi_evolution} shows that $\langle
\Nsat\rangle$ is {\rm higher} at high $z$.  Instead, satellite
fractions are lower at high $z$ because all relevant host halos have
$\Mhost > \Mstar$ and lie on the  exponentially-decreasing portion of
the halo mass function, so the relative number of $\Mone$-mass host
halos to  $\Mmin$-mass halos decreases with redshift.  Nevertheless,
these satellite fractions at high redshift are too  high to support a
power-law galaxy correlation function.}
\item{Figure~\ref{fig:fixedNumDens} implies that the physical
mechanisms that dictate the HOD of galaxies operate to  maintain
$\alpha$ and $\fsat$ approximately fixed and {\em not} to achieve a
power-law correlation function.  }
\end{enumerate}
We have established that the observed power-law correlation function
at low masses and low redshifts should not  persist at higher masses
or redshifts for simple, physical reasons.  However, exploring the HOD
parameter space has not  revealed a single simple property that yields
a power-law shape for $\xir$.  In an effort to better understand the
factors that drive a power-law $\xir$ at high precision, we continue
to explore the HOD parameter space in a different  way.  Specifically,
we investigate the two mass scales in the standard HOD models, $\Mmin$
and $\Mone$,  relative to the characteristic non-linear collapse mass,
$\Mstar$.  To complement our previous analysis and to  be consistent
with gross theoretical predictions, we fix $\alpha=1$ and take our two
parameters to be $\Mmin/\Mstar$ and  $\Mone/\Mmin$.  The first ratio
specifies roughly the host masses that galaxies occupy relative to the
exponential  regime of the halo mass function, and the second ratio
sets the length of the ``plateau'' in the HOD.

\begin{figure}
\begin{center}
\includegraphics[height=.75\textheight,width=.45\textwidth,angle=0]{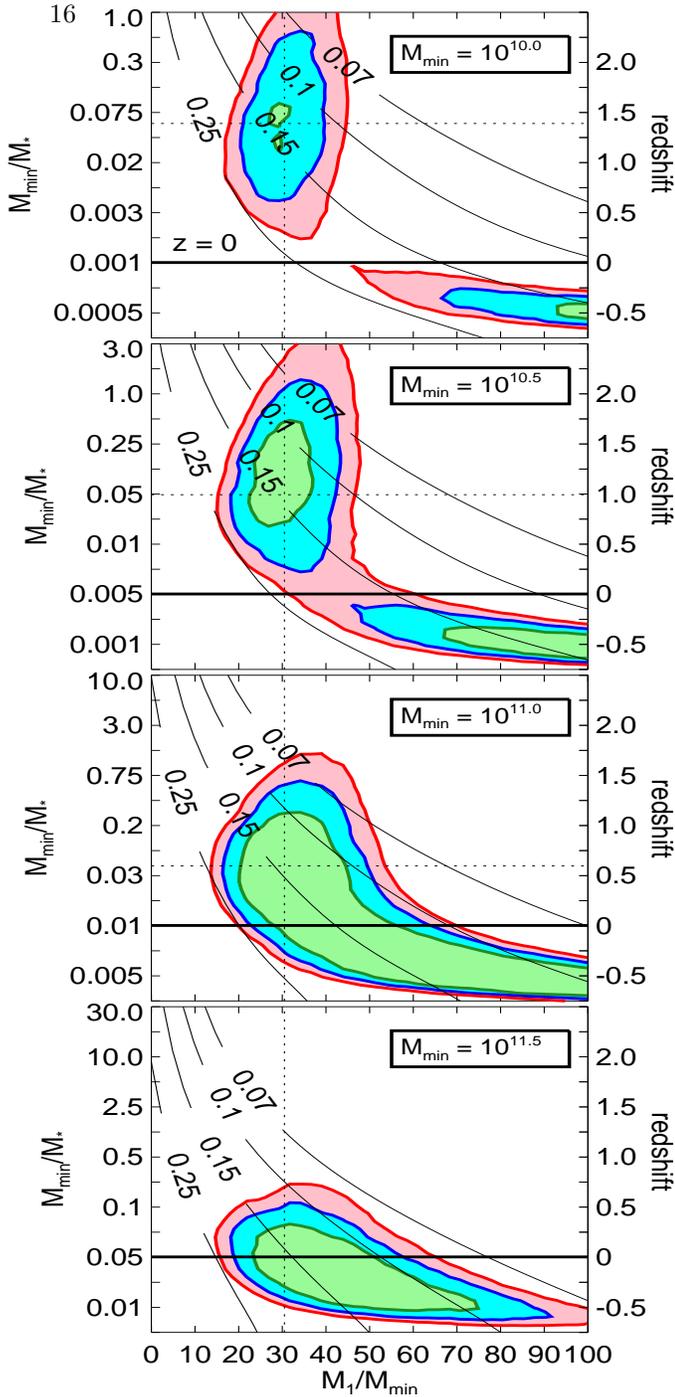}
\caption{Exploration of the HOD parameter space that yields a
power-law $\xir$.  Each panel corresponds to a different  mass
threshold $\Mmin$ (in units of $\hMsun$).  The y-axis shows redshift
(right-hand side), which also corresponds  directly to the ratio
$\Mmin / \Mstar$ (left-hand side), since the characteristic non-linear
mass $\Mstar$ depends  directly on redshift.  The horizontal axis
shows the ratio $\Mone / \Mmin$.  We fix the slope of the satellite
mean  occupation function to be $\alpha=1$ and we set the fourth HOD
parameter $\Mzero$ using equation~\ref{eqn:Mzero}.  Each  point on the
horizontal axis therefore corresponds to a specific set of HOD
parameters, while moving along the vertical  axis shifts the HOD to
different redshifts.  As in Fig.~\ref{fig:fixedNumDens}, shaded
contours represent the  $68.3\%, 95\%$ and $99.6\%$ power-law
likelihood spaces and thin solid curves show contours of constant
satellite  fraction.  The horizontal and vertical dotted lines
correspond to fixed values of $\Mmin / \Mstar = 0.05$ and  $\Mone /
\Mmin = 30$, which bisect the best-fit power-law space in all four
panels.  Solid black horizontal lines  denote $z = 0$, below which the
parameter space corresponds to future epochs.}
\label{fig:chisqr_mratios}
\end{center}
\end{figure}

Figure~\ref{fig:chisqr_mratios} probes the power-law $\xir$ space as a
function of the ratios $\Mmin/\Mstar$ and  $\Mone/\Mmin$.  For this
analysis we switch from fixed number density samples to fixed mass
thresholds, and we choose  four values of minimum mass that correspond
to a range of sub-$\Lstar$ galaxies ($\Mmin = 10^{10.0}, 10^{10.5},
10^{11.0},  10^{11.5} \hMsun$), showing results for each in a distinct
panel.  In each panel, we sample redshifts from $z=-0.9$ to  $z=2.9$
in steps of $\Delta z=0.08$ (labeled on the right vertical axis).
Each redshift value also corresponds to a  $\Mmin / \Mstar$ ratio,
which we label on the left vertical axis.  At each redshift, we also
loop over $\Mone / \Mmin$  ratios from 1 to 100 in steps of $\Delta
(\Mone/\Mmin)=2$.  For every pair of $\Mmin / \Mstar$ and $\Mone /
\Mmin$  values, we compute $\xir$ using the halo model and fit a
power-law function in the same fashion as described previously.   As
before, we show the $68.3\%, 95\%$ and $99.6\%$ likelihood regions of
$\xir$ consistent with a power law (green,  blue, and red contours,
respectively).  Also, as before, we show contours of constant
satellite fraction, $\fsat$ (solid  black curves).  The thick
horizontal lines at $z=0$ are meant to emphasize that the parameter
space lying below these  lines corresponds to {\em future} epochs.

We have repeated this analysis for higher mass thresholds (values of
$\Mmin = 10^{12.0}, 10^{12.5}, 10^{13.0},  10^{13.5} \hMsun$).
However, we do not show those results because we found no parameter
combinations within the $99.6\%$  power-law likelihood space.
Figure~\ref{fig:fixedNumDens} showed that the power-law parameter
space drifted to higher  values of $\alpha$ for lower number density
(and hence higher mass) samples in an effort to drive up the two-halo
term  to meet the enhanced one-halo term.  Therefore, it is not
surprising that we do not find this space when  we restrict the slope
to be $\alpha=1$.

Many interesting results can be drawn from
Figure~\ref{fig:chisqr_mratios}.   Again, we itemize them for the sake
of  clarity.
\begin{enumerate}
\label{chisqr_mratios_summary}
\item{In order for $\xir$ to have a shape consistent with a power-law
assuming $\sim 10\%$ measurement errors, it  appears necessary for the
``plateau'' in the HOD to be sufficiently long.  At all masses and
redshifts, $\Mone/\Mmin$  needs to be at least $\sim 20$, otherwise
the large satellite fraction drives a one-halo term that is too large
relative  to the two-halo term.  Moreover, for past epochs, $z>0$, the
maximum plateau length is $\Mone/\Mmin \lesssim 40$.   Higher values
of $\Mone/\Mmin$ yield a one-halo term that is too weak.  In fact,
$\Mone/\Mmin \sim 30$ seems to be the  preferred value to yield a
nearly power-law correlation function at all masses so long as $z>0$.
This value is denoted  by the vertical dotted lines in all the panels.}
\item{For $z \geq 0$, a near power-law $\xir$ seems to require a
restricted range of $\Mmin/\Mstar$.  Interestingly, the  value
$\Mmin/\Mstar \sim 0.05$ can yield a power-law correlation function at
all masses for appropriate choices of  redshift.  This value is
denoted by the horizontal dotted lines in all the panels.  This
restriction on  $\Mmin/\Mstar$ means that higher redshift samples
(when $\Mstar$ is significantly smaller than today) can  only exhibit
power-law behavior if the relevant host halos are significantly
smaller.  This possibility becomes  irrelevant in a practical sense
because star formation is inefficient in small halos  \citep[$M \ll
10^{11} \hMsun$, e.g.,][]{conroy_wechsler09,behroozi10,guo10}), so
they  cannot host galaxies that are easily observable at high
redshift.  Figure~\ref{fig:chisqr_mratios} shows that the  lowest-mass
samples that we consider (top two panels) have a nearly power-law
$\xir$ at $1 \lesssim z \lesssim 2$,  whereas the highest-mass samples
have a nearly power-law $\xir$ only at low redshift.}
\item{At sufficiently high redshift, near power-law clustering is no
longer achievable at any  mass threshold corresponding to relatively
bright galaxies.  Our results generally indicate that  power-law
clustering at high redshift can only be achieved if galaxies at high
redshift occupy halos in a  markedly different and more complicated
manner than their low-$z$ counterparts.}
\item{For future epochs these broad results no longer hold.  A broader
range of $\Mone/\Mmin$ values can be made  approximately consistent
with a power law at low values of $\Mmin/\Mstar$, or low/negative
redshifts.  For the lowest  $\Mmin$ samples the power-law likelihood
space is clearly bimodal, with possible ways to achieve a power law
both at  high redshifts and at low/future redshifts.}
\item{At all masses and redshifts we find that the power-law
likelihood parameter space has satellite fractions in the  range
$\fsat \sim 0.1-0.25$, with the $\fsat=0.15$ contour slicing through
all of the $1\sigma$ regions.  $\fsat$ is  naturally strongly
dependent on both $\Mone/\Mmin$ and $\Mmin/\Mstar$.  Increasing the
length of the HOD plateau at  fixed $\Mmin$ and redshift makes $\fsat$
decrease, as does boosting $\Mmin/\Mstar$ while keeping $\Mmin$ and
the  plateau fixed.  If we keep both ratios fixed (i.e., both the HOD
shape and its position relative to the mass function)  then the
satellite fraction is also approximately fixed, regardless of $\Mmin$.}
\end{enumerate}
%

\section{DISCUSSION AND PRIMARY CONCLUSIONS}
\label{summary}

It has been recognized for decades that the two-point correlation
function has a simple, power-law form with  $\xir \sim r^{-2}$.
Observational determinations of galaxy two-point clustering spanning
more than thirty years all  yielded results consistent with a single
power law extending from linear and quasi-linear length scales  ($r
\gtrsim 30\, \hmpc$) to deeply non-linear scales ($r \lesssim 0.1\,
\hmpc$).  In this paper, we cast the  problem in the contemporary
setting in which galaxies form in halos and subhalos of dark matter
and set out to understand the  physical processes that drive this
surprisingly simple result.  Our primary conclusion is that the nearly
power-law  correlation function of relatively common, $\Lstar$ and
sub-$\Lstar$ Galaxies at $z \sim 0$ is a coincidence and does  not
reflect any general principle of structure formation or galaxy
evolution. So how did we  arrive at this conclusion?

First, the efficiency of galaxy formation is dependent upon halo mass
and it has been determined both theoretically  and empirically that
there is a halo mass scale below which galaxy formation is
inefficient, roughly  $M_{\mathrm{gal}} \sim 10^{10.5}\, \hMsun$
\citep{conroy_wechsler09,behroozi10,guo10}.  A number of things can
set this  scale including atomic and molecular physics and feedback
from supernovae and active galactic nuclei \citep[for a  recent review
article see][]{BensonReview10}.  This mass scale is $M_{\mathrm{gal}}
< \Mstar$, so $\Lstar$ and  sub-$\Lstar$ galaxies are common.  Had
$M_{\mathrm{gal}}$ been greater than or similar to $\Mstar$, most
bright galaxies  would lie in comparably rare halos and be rare
themselves.  In such a case, one-halo clustering would be  too strong
to be compatible with a power law.  $\Mstar$ is {\em not} determined
by galaxy formation physics but  is set by the completely unrelated
processes that establish the amplitude of cosmological density
fluctuations, presumably primordial inflation.

Second, power-law clustering requires that some of the galaxies formed
within relatively large subhalos are destroyed.  Destruction
is due primarily to mass loss, and, to a lesser extent, merging with
the central galaxy as a result of dynamical friction.  Without this
destruction, satellite fractions would be too high and small-scale
clustering  too strong compared with large-scale clustering.  In a
forthcoming paper, we perform more sophisticated modeling to make the
connection between subhalo mass loss and stellar mass loss in order to
make predictions for the amount of intracluster light.  Large-scale
clustering is principally set by large-scale matter  density
fluctuations and is insensitive to the details of galaxy formation
within halos, while the strength of small scale clustering grows in
proportion to the fraction of galaxies that are satellites and in
inverse proportion to the number density of the galaxies of interest.
As it turns out, precisely the right amount of subhalo destruction has
occurred by redshift $z \sim 0$ in a concordance cosmology to produce
a single, unbroken, power-law $\xir$.

Evolution of the satellite fraction is set by a competition between
halo mergers, which increase $\fsat$, and  destruction by dynamical
processes, which occur on a dynamical timescale and reduce $\fsat$.
At high redshifts, mergers  occur more rapidly than destruction for
halos with masses $\gtrsim M_{\mathrm{gal}}$.  The low-redshift merger
rate  declines in part due to the fact that $M_{\mathrm{gal}} <
\Mstar$ at $z \lesssim 1$.  Halos with masses  below $\Mstar$ become
relatively more likely to merge with a larger object than to acquire
new substructure  compared to counterparts with masses greater than
$\Mstar$ \citep[see][]{zentner07}.  More importantly,  the rate of
halo mergers is quenched at $z \lesssim 1$ as dark energy begins to
suppress further  cosmological structure growth.  As merger rates
decline, satellites are depleted with time.  Therefore, at  $z \sim
0$, the correlation function is nearly a power law because the
competition between the accretion and  destruction rates has struck
just the right balance to yield the appropriate value of $\fsat$.

The merger and destruction rates will once again become unbalanced in
the future as halo merging  is stifled by dark energy and existing
satellite galaxies are slowly destroyed over many dynamical times
through complex interactions in their host environments.  We show that
this will result   in small-scale clustering that will be
significantly {\em too weak} to be consistent with a power law.

Largely as a consequence of the merger/destruction competition, $\xir$
evolves through cosmic time, achieving a power law only near $z\sim 0$
for  $L \sim \Lstar$ and dimmer galaxies.  The  processes of galaxy
formation, the amplitude of cosmological density fluctuations, the
abundance of  dark matter, and the nature of the dark energy are
thought to be completely distinct and determined by  {\em unrelated
physics}.  So the power-law $\xir$ at $z \sim 0$ is a coincidental
conspiracy.

In establishing these broad conclusions, we have performed an
exhaustive investigation of the ingredients of the galaxy  correlation
function, which has revealed many interesting, more detailed
conclusions.  These can be summarized as follows.
\begin{enumerate}
\label{main_results}
\item{We find that satellite halo mass loss is the principle dynamical
process responsible for depleting sufficient  substructure so as to
nearly align the one- and two-halo terms to yield a power-law
correlation function at low  redshift.  Dynamical friction plays a
smaller supporting role, accounting for an additional  $\sim 15\%$ of
subhalo destruction.}

\item{The shape of the correlation function is strongly mass
dependent.  For instance, at low redshift deviations from  a power law
$\xir$ grow with increasing host halo mass. This drives stronger
deviations from a power law for higher  luminosity galaxy samples.
The best power-law fits derived from our model are for galaxies
residing in halos that are  common enough to correspond to $\sim
\Lstar$ and dimmer galaxies, in agreement with observations.}

\item{The correlation function is highly redshift-dependent.  The
sensitivity of the one-halo term to the HOD,  coupled with the
relative insensitivity of the two-halo term, implies that achieving a
power-law requires fine-tuning  the number of satellite galaxies per
halo.  The satellite galaxy abundance evolves with redshift, driven by
the evolving  balance between accretion and destruction, with an
enhanced amount of substructure at high redshift.  Therefore, the
correlation function can only achieve a power law during those epochs
when substructure has evolved to align the one-  and two-halo terms.
The correlation function is boosted on small scales at high $z$, the
one- and two-halo terms  join at $z = 0$ to form a power-law, then the
power law is once again broken in future epochs.}

\item{For three chosen number densities corresponding to low-redshift,
$\sim \Lstar$ and dimmer galaxies, we probed the  most likely
power-law space as a function of redshift for a parametrized HOD.  We
find that there is a relatively narrow  range of satellite fractions
for $\xir$ to be consistent with a single power law (assuming $\sim
10\%$ measurement errors)  at any given redshift.  At all redshifts
and masses, power-law correlation functions have satellite fractions
in the  range $\fsat \sim 0.1 - 0.25$.  It is difficult to achieve a
power-law correlation function at $z \gtrsim 3$  for any number
density.}

\item{We find that to achieve a power law $\xir$ at high mass or
redshift, the slope $\alpha$ of the satellite galaxy occupation
function must be significantly steeper than unity (for instance,
greater than 2 at $z=3$).  This would imply that the mapping of
galaxies to halos is much more complicated than we  think, since the
number of galaxies would have to be very different than  the number of
subhalos of a particular size.  Instead, it appears that the processes
that govern galaxy formation do not care about the conditions needed
to achieve a power law $\xir$.  }
\item{The ratio $\Mone/\Mmin$ (the ``plateau'' of the HOD) is a key
ingredient for predicting the shape of $\xir$.  The  prominence of the
plateau is a measure of substructure abundance.  Along with
$\Mone/\Mmin$, it is also necessary to  characterize the ratio
$\Mmin/\Mstar$, which specifies what halo masses galaxies occupy
relative to the halo mass  function.  By maintaining the combination
of $\Mone/\Mmin \sim 30$ and $\Mmin/\Mstar \sim 0.05$ we can achieve a
near  power law for redshifts in the range $0-1.5$ and the appropriate
mass threshold at each redshift (the mass threshold  is $\Mmin \sim
\Mstar/20$, with $\Mstar$ set by the redshift).  At higher redshifts
this criterion is met for galaxies  that are most likely too dim to be
observed.  For example, achieving the requisite $\Mmin \sim \Mstar/20$
at $z = 2$  corresponds to a halo mass of $\Mmin \sim 10^{9}\, \hMsun$
in which star formation is inefficient.  }
\end{enumerate}

This work has allowed us to formulate a general picture of the nature
of the galaxy two-point correlation function.  Halo abundances and
subhalo populations evolve with time.  At high redshifts, halos large
enough to harbor galaxies  are rare and subhalos are abundant within
these hosts.  With time, host halos that harbor galaxies generally
become  more common (though the specifics of this evolution can be
subtle) and subhalos within these hosts become relatively  less
abundant.  All the while, large-scale matter correlations grow, but
the clustering bias of large halos  evolves to largely compensate for
this large-scale growth of structure.   These effects, considered
either individually or in tandem, change the HOD and the shape of
$\xir$.   As a result, the correlation function evolves through an
epoch where it is close to a power law and this epoch happens to be
near $z\sim 0$.  From our broad discussion and detailed conclusions,
it is clear that a nearly power-law  correlation function requires a
conspiracy between otherwise unrelated processes such as the  early
Universe physics that established the initial conditions for low
redshift structure,  the detailed physical processes that determine
galaxy and star formation efficiency, and  the growth rate of cosmic
structure set largely by the abundances of dark matter and dark
energy.   The low-redshift power-law galaxy two-point function is thus
a mere cosmic coincidence.

\acknowledgments

We appreciate many helpful discussions from James Bullock, Scott
Drake, Andrey Kravtsov, Cameron McBride, Jeff Newman,  Risa Wechsler,
and Zheng Zheng.  AAB is supported by Vanderbilt University and the
Alfred P. Sloan Foundation.   The work of ARZ is funded by the
University of Pittsburgh and by the National Science Foundation
through  grant AST 0806367.

\bibliography{/data2/dwatson/lib/bibliography/citations}

\end{document}